\documentclass{article}

\usepackage{fullpage}

\newcommand{\R}{\proglang{R}}
\usepackage{algorithm}
\usepackage{algorithmic}

\usepackage{hyperref}
\usepackage{xcolor}
\definecolor{Ckt}{HTML}{E41A1C}
\definecolor{Min}{HTML}{4D4D4D}
\definecolor{MinMore}{HTML}{377EB8}
\definecolor{Data}{HTML}{984EA3}

\usepackage{tikz}
\usetikzlibrary{arrows}
\usepackage{amssymb,amsmath}
\usepackage{natbib}
\usepackage{amsthm}

\usepackage{stfloats}

\DeclareMathOperator*{\minimize}{minimize}
\newcommand{\ZZ}{\mathbb Z}

\newcommand{\RR}{\mathbb R}

\usepackage{thumbpdf,lmodern}

\usepackage{framed}

\newcommand{\proglang}[1]{#1}
\newcommand{\pkg}[1]{\textbf{#1}}
\newcommand{\code}[1]{\texttt{#1}}

\author{Toby Dylan Hocking, \code{Toby.Hocking@nau.edu} \\
    Guillem Rigaill, \code{Guillem.Rigaill@inra.fr}\\
    Paul Fearnhead, \code{p.fearnhead@lancaster.ac.uk}\\
    Guillaume Bourque, \code{guil.bourque@mcgill.ca}
 }

\title{Generalized Functional Pruning Optimal Partitioning (GFPOP)
  for Constrained Changepoint Detection in Genomic Data}

\usepackage{Sweave}
\begin{document}

\maketitle

\begin{abstract}
We describe a new algorithm and \proglang{R} package for
  peak detection in genomic data sets using constrained
  changepoint algorithms. These detect changes from background to peak
  regions by imposing the constraint that the mean should alternately
  increase then decrease. An existing algorithm for this problem
  exists, and gives state-of-the-art accuracy results, but it is
  computationally expensive when the number of changes is large. We
  propose the GFPOP algorithm that jointly estimates the number of
  peaks and their locations by minimizing a cost function which
  consists of a data fitting term and a penalty for each
  changepoint. Empirically this algorithm has a cost that is
  $O(N \log(N))$ for analysing data of length $N$. We also propose a
  sequential search algorithm that finds the best solution with $K$
  segments in $O(\log(K) N\log(N))$ time, which is much faster than
  the previous $O(K N\log(N))$ algorithm. We show that our disk-based
  implementation in the \pkg{PeakSegDisk} \R\ package can be used
  to quickly compute constrained optimal models with many changepoints,
  which are needed to analyze typical genomic data sets
  that have tens of millions of observations.
\end{abstract}

\section[Introduction]{Introduction} \label{sec:intro}

\subsection{Peak detection via changepoint methods}

There are many applications, particularly within genomics, that
involve detecting regions that deviate from a usual/background
behaviour, and where qualitatively these deviations lead to an
increased mean of some measured signal. For example, ChIP-seq data
measure transcription factor binding or histone modification
\citep{chip-seq}; ATAC-seq data measure open chromatin \citep{ATACseq}. In these data
we have counts of aligned reads at different positions along a
chromosome, and we would like to detect regions for which the count data are larger than the usual background level.

One approach to detecting these regions is through algorithms that detect changes in the mean of the data. This paper
builds on recent work of \citet{Hocking-constrained-changepoint-detection} and presents a new changepoint algorithm, and its implementation in
\R. This algorithm is based on modeling count data using a Poisson distribution, and using the knowledge that we have
background regions with small values and peak regions with large values. This imposes constraints
on the directions of changes, with the mean of the data alternately increasing then decreasing in value. A particular
challenge with genomic data is that for an algorithm to be widely used, it must scale well  to large data in terms of 
both time and memory costs.

There are other algorithms for tackling this type of problem, for
example based on hidden Markov models \citep{chipseq-hmm}.  One drawback of such
methods is that they assume the background/peak means do not change across
large genomic regions, whereas such long-range changes are observed in
many real data sets. For a detailed comparison of other algorithms with
changepoint approaches we refer the reader to
\citep{HOCKING2016-chipseq}; we focus the remainder of the paper on
optimal changepoint models.

\subsection{Optimal changepoint models with no constraints between adjacent segment means}

Denote the data by $z_1,\dots,z_N$. We assume the data is ordered: for genomic applications the ordering will 
be due to position along a chromosome, for time-series data the ordering is commonly by time. The aim of 
changepoint analysis is to partition the data in to $K$ segments that each contain consecutive data points, 
such that features of the data are common within a segment but differ between segments. The feature of the data 
that changes will depend on the application, but could be, for example, the mean of the data, the variance, or 
the distribution. Detecting changes of different features requires different statistical algorithms.

Throughout we will let $K$ be the number of segments, with the changepoints being $0=t_0<t_1<\cdots<t_{K-1}<t_K=N$. 
This means that the $k$th segment will contain data points $z_{t_{k-1}+1},\ldots,z_{t_k}$. We denote the 
segment-specific parameter for the segment by $m_k$. For the problem of detecting changes in ChIP-seq count data,
the simplest statistical model uses Poisson random variables with segment-specific mean  
parameters for that segment. 
Change detection is then an attempt to detect the points along the 
chromosome where the mean of the data changes. 

The algorithm we present is based on detecting changes via minimizing
a measure of fit to the data, with this measure of fit being the
negative log-likelihood under our Poisson model. This corresponds to
using the loss function $\ell(m,z)=m-z\log m$ for fitting a non-negative count data point
$z\in\ZZ_+$ with a mean parameter $m\in\RR_+$. If we know the number of
segments $K$ we can estimate the location of the segments by solving the
following minimization problem,
 \begin{align}
    \minimize_{\substack{
  \mathbf m\in\RR^{K}
\\
   0=t_0<t_1<\cdots<t_{K-1}<t_K=N
  }} &\ \ 
    \sum_{k=1}^K\  \sum_{i=t_{k-1}+1}^{t_k} \ell(m_k,z_i).
\label{min:ut}
\end{align}
Optimizing by naively searching over all possible arrangements of changepoints is an expensive $O(N^K)$ time
operation. However, solving (\ref{min:ut}) can be achieved efficiently using dynamic programming. The first such
algorithm was the Segment Neighborhood algorithm, which computes the series of 
optimal segmentations with  1 to $K$ segments in $O(KN^2)$ time
\citep{segment-neighborhood}.  The classical algorithm for solving the
Segment Neighborhood problem is available in \R\ as
\code{changepoint::cpt.mean}. Recent research has led to faster algorithms, based on pruning the search space of 
the Segment Neighborhood algorithm
\citep{pruned-dp-new,johnson}, and these algorithms empirically take $O(K N \log N)$ time. 
The novelty of these techniques is a functional
representation of the optimal cost, which allows pruning of the $O(N)$
possible changepoints to only $O(\log N)$ candidates (while
maintaining optimality). The original implementation of the PDPA was
available in \R\ as \verb|cghseg:::segmeanCO| for the Normal
homoscedastic model, but \pkg{cghseg} has been removed from CRAN as of
18 December 2017. The PDPA for the Normal homoscedastic model is now
available as \verb|jointseg::Fpsn| on Bioconductor
\citep{jointseg}. \citet{cleynen2013segmentation} described a
generalization of the PDPA for other likelihood/loss functions
(Poisson, negative binomial, Normal heteroscedastic). These are
available in \R\ as \verb|Segmentor3IsBack::Segmentor|.

In practice it is unusual to know how many segments there are present in the data. 
To estimate $K$ it is common to use some information criteria that takes account 
both of the measure of fit to the data and the complexity of the segmentation
model being fitted. The most natural measures of complexity are linear in the 
number of changepoints. Whilst it is possible to estimate $K$ by solving the 
Segment Neighborhood problem for an appropriate set of changes, and calculating 
the value of the information criteria for each value of the number of segments, it 
is faster to jointly estimate both $K$ and the changepoint locations that minimise 
the information criteria. The first algorithm to do so was the Optimal Partioning 
algorithm introduced by \citet{optimal-partitioning}. Optimal partitioning is an  
$O(N^2)$ algorithm, and can be significantly faster than Segment Neighborhood for 
large $K$. 

There has also been substantial research into speeding up the optimal partitioning algorithm, 
using various ideas to prune the search space. In particular the Pruned Exact Linear Time (PELT)
algorithm of \citet{pelt}, which is implemented within \code{changepoint::cpt.mean}, and Functional Pruning
Optimal Partitioning (FPOP) of \citet{Maidstone2016} which is available in \R\ as
\verb|fpop::Fpop| \citep{fpop}. These algorithms have a computational cost of 
$O(N)$ if $K$ increases linearly with $N$. The FPOP algorithm has a computational cost 
that is empirically $O(N \log N)$ in situations where $K$ increases sub-linearly with $N$. 
See Table \ref{tab:unconstrained-algos} for a summary of the different dynamic programming algorithms and implementations.

Whilst solving the optimal partitioning problem is faster than solving (\ref{min:ut}) for a range of $K$,
the drawback is that you only get a single segmentation for a single value $K$ of the number of segments. 
Furthermore the choice of penalty that you impose with the information criteria -- which corresponds the 
improvement in fit to the data needed to add an additional changepoint -- can be hard to tune and have an 
important impact on the accuracy of the estimate of the number of changepoints. One way to ameliorate this 
concern is to find segmentations for a range of penalties, which can be done efficiently \citep{crops}.

\begin{table*}[t!]
  \centering
  \begin{tabular}{r|c|c}
    Problem & No changepoint pruning & Functional pruning \\
    \hline
    Segment  & Dynamic Prog. Algo. (DPA) & Pruned DPA (PDPA) \\
                  Neighborhood & Optimal, $O(KN^2)$ time & Optimal, $O(KN\log N)$ time\\
            $K$ segments& \citet{segment-neighborhood} & \citet{pruned-dp-new}\\
            & \pkg{changepoint} & \pkg{jointseg}\\
    \hline
    Optimal  & Optimal Partitioning Algorithm & FPOP \\
    Partitioning & Optimal, $O(N^2)$ time & Optimal, $O(N\log N)$ time\\
            Penalty $\lambda$& \citet{optimal-partitioning} & \citet{Maidstone2016}  \\
    &  & \pkg{fpop}\\
    \hline
  \end{tabular}
  \caption{Previous work on algorithms for optimal changepoint detection with 
    no constraints between adjacent segment means.}
\label{tab:unconstrained-algos} 
\end{table*}

There are alternative approaches to fitting changepoint models, the most common of which are based on 
specifying a test for a single change and then repeatedly applying this test to identify multiple 
changepoints. Such approaches can be applied more
widely than the dynamic programming based approaches described above, 
and often have strong computational performance with algorithms that are 
$O(N \log N)$ for the Segment Neighborhood problem. In situations where both procedures 
can be used, these methods are often identical if we wish to identify at most one changepoint. 
The advantage that the dynamic programming approaches have is that they jointly detect multiple 
changepoints which can lead to more accurate estimates \cite[see e.g.][]{Maidstone2016}. Several of 
these alternative algorithms are available in \R.   For example, the \pkg{wbs} package implements
the wild binary segmentation method of \citet{wbs}. An efficient
implementation of the classical binary segmentation heuristic is
available as \code{fpop::multiBinSeg}. The
\pkg{stepR} package implements the SMUCE algorithm for multiscale
changepoint inference \citep{stepR}.

\subsection{Models with inequality constraints between adjacent segment means}

The models discussed above are unconstrained in the sense that there
are no constraints between mean parameters $m_k$ on different
segments. However, as described above, constraints can be useful when
data need to be interpreted in terms of pre-defined domain-specific
states. In the ChIP-seq application the changepoint model needs to be
interpreted in terms of peaks (large values which represent protein
binding/modification) and background (small values which represent noise).

In this context, \citet{HOCKING-PeakSeg} introduced a $O(KN^2)$
Constrained Dynamic Programming Algorithm (CDPA) for fitting a model
where up changes are followed by down changes, and vice versa
(Table~\ref{tab:constrained-algos}). These constraints ensure that
odd-numbered segments can be interpreted as background, and
even-numbered segments can be interpreted as peaks. Although the CDPA
provides a sub-optimal solution to the Segment Neighborhood problem in
$O(KN^2)$ time, \citet{HOCKING2016-chipseq} showed that it achieves
state-of-the-art peak detection accuracy in a benchmark of
ChIP-seq data sets. 

Because the quadratic time complexity of the CDPA limits its
application to relatively small data sets,
\citet{Hocking-constrained-changepoint-detection} proposed to generalize
the functional pruning method for changepoint models with
constraints between adjacent segment means. The resulting Generalized
Pruned Dynamic Programming Algorithm (GPDPA) reduces the number of
candidate changepoints from $O(N)$ to $O(\log N)$ while enforcing the constraints and maintaining
optimality. The GPDPA computes the optimal solution to the up-down
constrained Segment Neighborhood problem in $O(KN\log N)$ time.  The
\pkg{PeakSegOptimal} \R\ package provides an in-memory solver for the up-down constrained Segment Neighborhood model
\citep{Hocking-constrained-changepoint-detection}.

\begin{table}
  \centering
  \begin{tabular}{r|c|c}
    & No changepoint pruning & Functional pruning \\
    \hline
    Segment Neighborhood & Constrained DPA & Generalized PDPA \\
$K$ segments    & Sub-optimal,  $O(KN^2)$ & Optimal, $O(KN\log N)$\\
    & \citet{HOCKING-PeakSeg} & 
\citet{Hocking-constrained-changepoint-detection}
 \\
& \pkg{PeakSegDP} & \pkg{PeakSegOptimal}\\
    \hline
    Optimal Partitioning &  & Generalized FPOP \\
    Penalty $\lambda$&  & Optimal, $O(N\log N)$\\
    &  & \textbf{This work}\\
    & & \pkg{PeakSegDisk}\\
    \hline
  \end{tabular}
  \caption{Algorithms for optimal changepoint detection with up-down 
constraints on adjacent segment means. Previous work is limited to solvers for the Segment Neighborhood problem; this paper presents Generalized Functional Pruning Optimal Partitioning (GFPOP), Algorithm~\ref{algo:GFPOP}.}
  \label{tab:constrained-algos}
\end{table}



\subsection{Contributions}

This paper presents two new algorithms for constrained optimal
changepoint detection (Section~\ref{sec:algorithms}), along with an
analysis of their empirical time/space complexity in a benchmark of
genomic data (Section~\ref{sec:results}). The algorithms are
implemented in the \R\ package \pkg{PeakSegDisk} on
GitHub.\footnote{ \url{https://github.com/tdhock/PeakSegDisk} }

First, we present a new algorithm for solving the Optimal Partitioning
problem with up-down constraints between adjacent segment means
(GFPOP, Algorithm~\ref{algo:GFPOP}). The fastest existing algorithm
for the up-down constrained changepoint model was the
$O(KN\log N)$ solver for the Segment Neighborhood problem
(Table~\ref{tab:constrained-algos}). In large genomic data sets, we
are only interested in models with many segments/changepoints, so it
is a waste of time and space to compute all models from 1 to $K$
segments using Segment Neighborhood algorithms. Our proposed GFPOP
algorithm solves the Optimal Partitioning problem, so yields one
optimal model with $K$ segments (without having to compute the models
from 1 to $K-1$ segments). We show that the empirical complexity of
our GFPOP implementation is $O(N\log N)$ time, $O(N\log N)$ space, and
$O(\log N)$ memory, which makes it possible to compute optimal models with many peaks
for typical genomic data sets on common laptop computers.

Although solving the Optimal Partitioning problem is faster by a
factor of $O(K)$, the user is unable to directly choose the number of
segments $K$. The user inputs a penalty $\lambda$, and gets one of the
optimal changepoint models as output. Thus, we also propose a
sequential search (Algorithm~\ref{algo:seq-search}) which computes the
optimal model for a specified number of segments $K$. It repeatedly
calls GFPOP to solve Optimal Partitioning with different penalties
$\lambda$, until it finds the maximum likelihood model with at most $K$
segments. We empirically show that the sequential search only requires
$O(\log K)$ evaluations of GFPOP. Overall the proposed algorithm is
thus $O( N \log(N)\log(K))$ time, $O(N\log N)$ disk, $O(\log N)$
memory. In an analysis of benchmark genomic data sets, we show that
this algorithm can compute an optimal model with $O(\sqrt N)>1000$ peaks
for $N=10^7$ data using only hours of compute time and gigabytes of
storage (which is much less than weeks/terabytes which would be
required for the Segment Neighborhood solver). 

\section{Statistical models and optimization problems} \label{sec:models}

\subsection{Unconstrained Optimal Partitioning problem}

Define our loss function to be the Poisson loss, $\ell(m,z)=m-z\log m$, and let $\lambda>0$ be a penalty for adding a changepoint. Then we can infer the number of segments and the location of the changes by solving
the Optimal Partitioning problem
\begin{align}
  \label{min:op}
  \minimize_{
  \substack{
  \mathbf m\in\RR^N
  }
  } &\ \ 
      \sum_{i=1}^N \ell(m_i, z_i) + \lambda \sum_{i=1}^{N-1} I(m_i\neq m_{i+1}).
\end{align}
The first term measures fit to the data, and the second term measures model complexity, which is proportional to the number of
changepoints. 
The non-negative penalty $\lambda\in\RR_+$ controls the tradeoff
between the two objectives (it is a tuning parameter that must be
fixed before solving the problem). Larger penalty $\lambda$ values
result in models with fewer changepoints/segments. The extreme penalty
values are $\lambda=0$ which yields $N$ segments ($N-1$ changepoints),
and $\lambda=\infty$ which yields 1 segment (0 changepoints).

Below we write an equivalent version of the Optimal Partitioning
problem, in terms of changepoint variables $c_i$ and state variables
$s_i$:
\begin{align}
  \label{min:op-c}
  \minimize_{
  \substack{
  \mathbf m\in\RR^N,\, \mathbf s\in\{0\}^N\\
  \mathbf c\in\{0,1\}^{N-1}\\
  }
  } &\ \ 
      \sum_{i=1}^N \ell(m_i, z_i) + \lambda \sum_{i=1}^{N-1} I(c_i=1) \\
  \text{subject to\ \ } &\ \text{no change: }c_i = 0 \Rightarrow m_i = m_{i+1}\text{ and } s_i=s_{i+1}, \nonumber\\
    &\ \text{change: }c_i = 1 \Rightarrow m_i \neq m_{i+1}\text{ and }(s_i,s_{i+1})=(0,0).\label{eq:change-constraint}
\end{align}
Note that the state $s_i$ and changepoint $c_i$ variables could be
eliminated from the optimization problem --- $s_i=0$ and 
$c_i=I(m_i\neq m_{i+1})$ for all $i$. We include them in problem~(\ref{min:op-c})
in order to show the relationship with the problem in the next
section, with constraints between adjacent segment means.

\citet{Hocking-constrained-changepoint-detection} proposed to use a
graph to represent a constrained changepoint model. The graph that
corresponds to problem~(\ref{min:op-c}) is shown in
Figure~\ref{fig:state-graph}, left. In such graphs, nodes represent
possible values of state variables $s_i$ and edges represent possible
changepoints $c_i\neq 0$. Each edge/changepoint corresponds to a
constraint such as (\ref{eq:change-constraint}).

\subsection{Optimal Partitioning problem with up-down constraints between
  adjacent segment means}

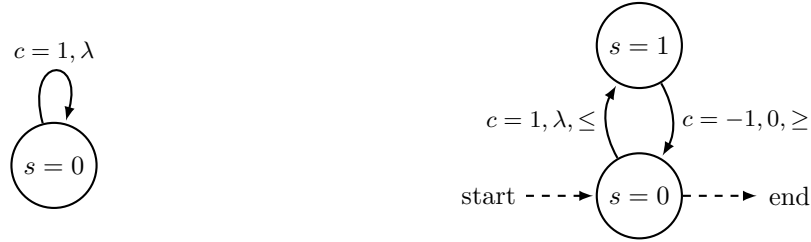
\begin{figure}
  \centering
  \begin{minipage}{3in}
    \centering
    \begin{tikzpicture}[->,>=latex,shorten >=1pt,auto,node distance=2cm,
      thick,main node/.style={circle,draw}]

      \node[main node] (0) {$s=0$};

      \path[every node/.style={font=\sffamily\small}]
      (0) edge [loop above] node {$c=1,\lambda$} (0);
    \end{tikzpicture}
  \end{minipage}
  \begin{minipage}{3in}
    \centering
    \begin{tikzpicture}[->,>=latex,shorten >=1pt,auto,node distance=2cm,
      thick,main node/.style={circle,draw}]

      \node[main node] (1) {$s=1$};
      \node[main node] (0) [below of=1] {$s=0$};
      \node (start) [left of=0] {start};
      \node (end) [right of=0] {end};

      \path[every node/.style={font=\sffamily\small}]
      (0) edge [bend left] node {$c=1, \lambda, \leq$} (1)
      (start) edge [dashed] (0)
      (0) edge [dashed] (end)
      (1) edge [bend left] node {$c=-1, 0, \geq$} (0);
    \end{tikzpicture}
  \end{minipage}
  \caption{State graphs for two changepoint models. Nodes represent
    states and solid edges represent changepoints. \textbf{Left:}
    one-state model with no constraints between adjacent segment
    means, problem~(\ref{min:op-c}). \textbf{Right}: two-state model with
    up-down constraints between adjacent segment means,
    problem~(\ref{min:op-up-down}). State $s=0$ represents background
    noise (small values) whereas state $s=1$ represents peaks (large
    values). Constraint $c=1$ enforces a non-decreasing change via the
    min-less operator ($\leq$) with a penalty of $\lambda$; $c=-1$
    enforces a non-increasing change via the min-more operator
    ($\geq$) with a penalty of $0$. The model is additionally
    constrained to start and end in the background noise $s=0$ state
    ($s_1=s_N=0$).}
  \label{fig:state-graph}
\end{figure}

For genomic data such as ChIP-seq \citep{chip-seq}, it is desirable to
have a changepoint model which is interpretable in terms of peaks
(large values) and background noise (small values). We therefore
propose a model based on the graph shown in
Figure~\ref{fig:state-graph}, right. It has two nodes/states: $s=0$
for background, and $s=1$ for peaks. It has two edges/changes: $c=1$
for a non-decreasing change from background $s=0$ to a peak $s=1$, and
$c=-1$ for a non-increasing change from a peak $s=1$ to background
$s=0$. Furthermore, the model is constrained to start and end in the
background state (because peaks are not present at the boundaries of
genomic data sequences). Maximum likelihood inference in
this model corresponds to the following minimization problem:
\begin{align}
  \label{min:op-up-down}
  F(\lambda) = \min_{
  \substack{
  \mathbf m\in\RR^N,\ \mathbf s\in\{0, 1\}^N\\
  \mathbf c\in\{-1, 0,1\}^{N-1}\\
  }
  } &\ \ 
      \sum_{i=1}^N \ell(m_i, z_i) + \lambda \sum_{i=1}^{N-1} I(c_i = 1) \\
  \text{subject to\ \ } &\ \text{no change: }c_i = 0 \Rightarrow m_i = m_{i+1}\text{ and }s_i=s_{i+1}, \nonumber\\
    &\ \text{non-decreasing change: }c_i = 1 \Rightarrow m_i \leq m_{i+1}\text{ and }(s_i,s_{i+1})=(0,1),\nonumber\\
    &\ \text{non-increasing change: } c_i = -1 \Rightarrow m_i \geq m_{i+1}\text{ and }(s_i,s_{i+1})=(1,0),\nonumber\\
  & \ \text{start and end down: } s_1=s_N=0.\nonumber
\end{align}
Note how the problem~(\ref{min:op-up-down}) with up-down constraints
is of the same form as the previous unconstrained
problem~(\ref{min:op-c}). Again there is one constraint for every
edge/changepoint in the state graph
(Figure~\ref{fig:state-graph}). The difference is that in
problem~(\ref{min:op-up-down}), we have inequality constraints between
adjacent segment means (e.g. when $c_i=1$, we must have a
non-decreasing change in the mean $m_i\leq m_{i+1}$). Another
difference is the model complexity in problem~(\ref{min:op-up-down})
is the total number of $c_i=1$ non-decreasing changes, which is
equivalent to the number of peak segments $P$, and is linear in the
total number of segments $K=2P+1$ and changes $K-1=2P$.

The solution to the Optimal Partitioning
problem~(\ref{min:op-up-down}) can be computed by first solving the
Segment Neighborhood version of the problem \citep{Maidstone2016}. In
\R\ the \pkg{PeakSegDP} package provides a sub-optimal solution in
$O(K N^2)$ time, and the \pkg{PeakSegOptimal} package provides an
optimal solution in $O(KN\log N)$ time. However in genomic data
the number of peaks/segments $K$ increases with $N$, so it is
intractable to solve the Segment Neighborhood problem because both $N$
and $K$ are large. Therefore in the next section we propose a new
algorithm for directly solving the constrained Optimal Partitioning
problem~(\ref{min:op-up-down}), which can yield a large number of
peaks in $O(N\log N)$ time.

\section{Algorithms and Software}
\label{sec:algorithms}
\subsection{Generalized Functional Pruning Optimal Partitioning (GFPOP)}

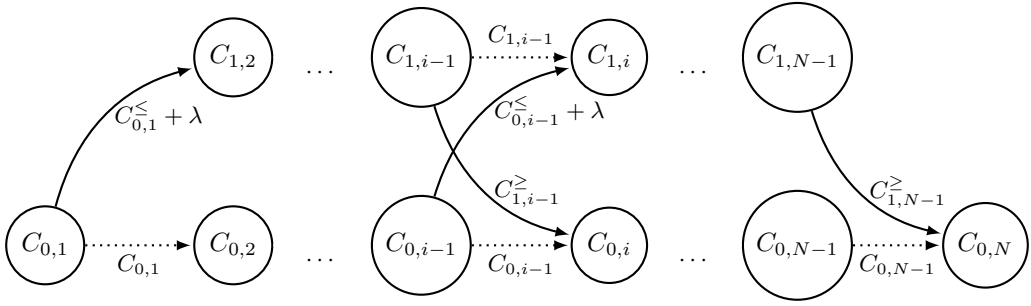
\begin{figure}
  \centering
  \begin{tikzpicture}[->,>=latex,shorten >=1pt,auto,node distance=2.5cm,
    thick,main node/.style={circle,draw}]
    \node[main node] (peak_t1) {$ C_{1,i-1}$};
    \node[main node] (bkg_t1) [below of=peak_t1] {$ C_{0,i-1}$};
    \node[main node] (peak_t) [right of=peak_t1] {$ C_{1,i}$};
    \node[main node] (bkg_t) [right of=bkg_t1] {$ C_{0,i}$};
    \node[main node] (peak_2) [left of=peak_t1] {$ C_{1,2}$};
    \node[main node] (bkg_2) [left of=bkg_t1] {$ C_{0,2}$};
    \node[main node] (peak_N1) [right of=peak_t] {$ C_{1,N-1}$};
    \node[main node] (bkg_N1) [right of=bkg_t] {$ C_{0,N-1}$};
    \node[main node] (bkg_N) [right of=bkg_N1] {$ C_{0,N}$};
    \node[main node] (bkg_1) [left of=bkg_2] {$ C_{0,1}$};
    \path[every node/.style={font=\small}]
    (peak_t1) edge [dotted] node {$ C_{1,i-1}$} (peak_t)
    (peak_t1) edge [black, bend right] node [right] {$ C_{1,i-1}^{\geq}$} (bkg_t)
    (bkg_t1) edge [dotted] node[midway, below] {$ C_{0,i-1}$} (bkg_t)
    (bkg_t1) edge [black, bend left] node[right] {$ C_{0,i-1}^{\leq}+\lambda$} (peak_t)
    (bkg_1) edge [black, bend left] node[right] {$ C_{0,1}^{\leq}+\lambda$} (peak_2)
    (bkg_1) edge [dotted] node[midway, below] {$ C_{0,1}$} (bkg_2)
    (peak_N1) edge [black, bend right] node [right] {$ C_{1,N-1}^{\geq}$} (bkg_N)
    (bkg_N1) edge [dotted] node[midway, below] {$ C_{0,N-1}$} (bkg_N)
    (bkg_2) edge [color=white] node[below, text=black, pos=0.5] {$\cdots$} (bkg_t1)
    (peak_2) edge [color=white] node[below, text=black, pos=0.5] {$\cdots$} (peak_t1)
    (bkg_N1) edge [color=white] node[below, text=black, pos=0.5] {$\cdots$} (bkg_t)
    (peak_N1) edge [color=white] node[below, text=black, pos=0.5] {$\cdots$} (peak_t)
    ;
  \end{tikzpicture}
  \caption{Directed acyclic graph (DAG) representing dynamic
    programming computations (Algorithm~\ref{algo:GFPOP}) for
    changepoint model with up-down constraints between adjacent
    segment means. Nodes in the graph repesent cost functions, and
    edges represent inputs to the the MinOfTwo sub-routine
    (solid=changepoint, dotted=no change). There is one column for
    each data point and one row for each state: the optimal cost of
    the peak state $s=1$ at data point $i$ is $ C_{1,i}$ (top
    row); the optimal cost of the background noise state $s=0$ is
    $ C_{0,i}$ (bottom row). There is only one edge going to
    $ C_{0,2}$ and $ C_{1,2}$ because the model is
    constrained to start in the background noise state ($s_1=0$).}
  \label{fig:computation-graph}
\end{figure}

In this section we propose a generalization of the FPOP algorithm
\citep{Maidstone2016} which allows optimal inference in models with
inequality constraints between adjacent means, such as
problem~(\ref{min:op-up-down}). In particular we implemented the
optimal changepoint model using the Poisson loss and the up-down
constraints. The state graph
(Figure~\ref{fig:state-graph}, right) can be converted into a directed
acyclic graph (Figure~\ref{fig:computation-graph}) that represents the
dynamic programming updates required to solve
problem~(\ref{min:op-up-down}). Each node in the computation graph
represents an optimal cost function, and each edge represents an input
to the $\min\{\}$ operation in the dynamic programming
equations~(\ref{eq:dp-over}) and (\ref{eq:dp-under}) below.

More precisely, we define the optimal cost of mean $\mu$ in state
$\sigma$ at any data point $\tau\in\{1,\dots,N\}$ to be
\begin{align}
  \label{eq:C_sigma_tau}
   C_{\sigma,\tau}(\mu)=\min_{
  \substack{
  \mathbf m\in\RR^\tau,\ \mathbf s\in\{0, 1\}^\tau\\
  \mathbf c\in\{-1, 0,1\}^{\tau-1}\\
  }
  } &\ \ 
      \sum_{i=1}^\tau \ell(m_i, z_i) + \lambda \sum_{i=1}^{\tau-1} I(c_i = 1) \\
  \text{subject to\ \ } &\ c_i = 0 \Rightarrow m_i = m_{i+1}\text{ and }s_i=s_{i+1}, \nonumber\\
    &\ c_i = 1 \Rightarrow m_i \leq m_{i+1}\text{ and }(s_i,s_{i+1})=(0,1),\nonumber\\
    &\ c_i = -1 \Rightarrow m_i \geq m_{i+1}\text{ and }(s_i,s_{i+1})=(1,0),\nonumber\\
  & \  s_1=s_N=0,\nonumber\\
& \ m_\tau=\mu,\, s_\tau=\sigma.\label{tau-constraints}
\end{align}
Note how the objective and constraints above are identical to the
up-down constrained Optimal Partitioning
problem~(\ref{min:op-up-down}) up to $\tau-1$ data points, but with
two added constraints at data point $\tau$ (\ref{tau-constraints}). At
data point $\tau$ the mean is constrained to be $m_\tau=\mu$ and the
state is constrained to be $s_\tau=\sigma$. The optimal cost
$C_{\sigma,\tau}(\mu)$ is a real-valued function that must be computed
by minimizing over all previous means $m_1,\dots,m_{\tau-1}$, states
$s_1,\dots,s_{\tau-1}$, and changes $c_1,\dots,c_{\tau-1}$.
It can be computed recursively using the dynamic programming
updates that we propose below.

The algorithm begins by initializing the optimal cost of the
background state at the first data point,
\begin{equation}
  \label{eq:init_1}
  C_{0,1}(\mu)=\ell(\mu, z_1).
\end{equation}
The computations for the second data point are also special, because
the model is constrained to start in the background state $s_1=0$. To
get to the background state $s_2=0$ at the second data point requires no
change ($c_1=0$), with a cost of
\begin{eqnarray}
  C_{0,2}(\mu)=C_{0,1}(\mu)+\ell(\mu, z_2).
\end{eqnarray}
Similarly, to get to the peak state $s_2=1$ at the second data point
requires a non-decreasing change ($c_1=1$), with a cost of
\begin{equation}
  \label{eq:C_12}
  C_{1,2}(\mu)=\min_{m_1\leq\mu} C_{0,1}(m_1)+\lambda + \ell(\mu, z_2)=C_{0,1}^\leq(\mu)+\lambda+\ell(\mu, z_2).
\end{equation}
Note that we were able to re-write the optimal cost function in terms
of a single variable $\mu$ by using the min-less operator,
\begin{equation}
  \label{eq:min-less}
  f^\leq(\mu) = \min_{x\leq\mu} f(x).
\end{equation}
The min-less operator was introduced by \citet{Hocking-constrained-changepoint-detection} in order to compute the optimal cost in the
functional pruning algorithm that solves the Segment Neighborhood
version of this problem. 

More generally, the dynamic programming update rules can be derived
from the computation graph (Figure~\ref{fig:computation-graph}). The
optimal cost of the peak state $s=1$ at data $i>2$ is
\begin{equation}
  \label{eq:dp-over}
   C_{1,i}(\mu) = \ell(\mu, z_i) + \min\{
   C_{1,i-1}(\mu),\, 
   C_{0,i-1}^\leq(\mu)+\lambda
\}.
\end{equation}
Note how the inputs to the $\min\{\}$ operation are the same as the
edges leading to the $C_{1,i}$ node in the computation graph
(Figure~\ref{fig:computation-graph}). 

Similarly, the optimal cost of
the background state $s=0$ is
\begin{equation}
  \label{eq:dp-under}
     C_{0,i}(\mu) = \ell(\mu, z_i) + \min\{
   C_{0,i-1}(\mu),\, 
   C_{1,i-1}^\geq(\mu)+\lambda
\},
\end{equation}
where the min-more operator is defined as
\begin{equation}
  \label{eq:min-more}
  f^\geq(\mu) = \min_{x\geq\mu} f(x).
\end{equation}
These dynamic programming computations are summarized in
Algorithm~\ref{algo:GFPOP}, Generalized Functional Pruning Optimal
Partitioning. The key to implementing the algorithm is to use a
PiecewiseFunction data structure that can exactly represent an
optimal cost function $C_{s,i}$. In the case of the Poisson loss,
each $C_{s,i}(\mu)$ is a piecewise function where each piece is of the
form $\alpha\mu + \beta\log\mu +
\gamma$. Therefore the optimal cost can be stored as a list
of intervals of $\mu\in[\text{MIN},\text{MAX}]$, each with
coefficients $\alpha,\beta,\gamma$.

\begin{algorithm}
\begin{algorithmic}[1]
\STATE Input: data set $\mathbf z\in\RR^N$, penalty constant $\lambda\geq 0$.
\STATE Output: vectors of optimal segment means $U\in\RR^{N}$ and ends $T\in\{1,\dots,N\}^{N}$
\STATE Initialize $2\times N$ empty PiecewiseFunction objects $C_{s,i}$ either in memory or on disk.
\label{line:initialize}
\STATE Compute min $\underline z$ and max $\overline z$ of $\mathbf z$.
\label{line:op-min-max}
\STATE
$ C_{0,1}\gets
 \text{OnePiece}(z_1, \underline z, \overline z)$
\label{line:first-cost}
\STATE for data point $i$ from 2 to $N$: // dynamic programming
\label{line:for-dp-t}
\begin{ALC@g}
  \STATE $ M_1\gets \lambda + \text{MinLess}(i-1,  C_{0,i-1})$ //cost of non-decreasing change
  \label{line:op-MinLess}
  \STATE $ C_{1,i}\gets \text{MinOfTwo}( M_1,  C_{1,i-1})+\text{OnePiece}(z_i, \underline z, \overline z)$
  \label{line:op-MinOfTwo}
  \STATE $ M_0\gets \text{MinMore}(i-1,  C_{1,i-1})$ //cost of non-increasing change
  \label{line:op-MinMore}
  \STATE $ C_{0,i}\gets \text{MinOfTwo}( M_0,  C_{0,i-1})+\text{OnePiece}(z_i, \underline z, \overline z)$
  \label{line:op-MinOfTwo-under}
\end{ALC@g}
\STATE $\text{mean},\text{prevEnd},\text{prevMean}\gets \text{ArgMin}( C_{0,n})$ // begin decoding
\label{line:op-ArgMin}
\STATE $\text{seg}\gets 1;\, U_{\text{seg}}\gets \text{mean};\, T_{\text{seg}}\gets \text{prevEnd}$
\label{line:op-store-i}
\STATE while $\text{prevEnd} > 0$:
\begin{ALC@g}
  \STATE if $\text{prevMean} < \infty$: $\text{mean}\gets \text{prevMean}$
  \STATE if $\text{seg}$ is odd: $\text{cost}\gets  C_{1,\text{prevEnd}}$ else $ C_{0,\text{prevEnd}}$
  \STATE $\text{prevEnd},\text{prevMean}\gets\text{FindMean}(\text{mean}, \text{cost})$
  \STATE $\text{seg}\gets \text{seg}+1;\, U_{\text{seg}}\gets \text{mean};\, T_{\text{seg}}\gets \text{prevEnd}$
\label{line:op-i+1}
\end{ALC@g}
\caption{\label{algo:GFPOP}Generalized Functional Pruning Optimal
  Partitioning (GFPOP) for changepoint model with up-down constraints
  between adjacent segment means.}
\end{algorithmic}
\end{algorithm}

\paragraph{Discussion of pseudocode.} Algorithm~\ref{algo:GFPOP} begins on line~\ref{line:initialize} by
initializing the array $C_{s, i}$ of optimal cost functions (either in
memory or on disk). It then computes the min $\underline z$ and max
$\overline z$ of the data (line~\ref{line:op-min-max}) and uses the
OnePiece sub-routine to initialize the optimal cost at the first data
point (line~\ref{line:first-cost}). Since the Poisson loss is
$\ell(\mu, z_1)=\mu-z_1\log \mu$, this first optimal cost function is
represented as the single function piece with interval/coefficients
$(\alpha=1,\beta=-z_1,\gamma=0,\text{MIN}=\underline
z,\text{MAX}=\overline z)$. 

The dynamic programming recursion in this algorithm is a loop over
data points $i$ (line~\ref{line:for-dp-t}). To compute $C_{1,i}$, the
penalty constant $\lambda$ is added to all of the result of MinLess
(line~\ref{line:op-MinLess}), before computing MinOfTwo and adding the
cost of the new data point (line~\ref{line:op-MinOfTwo}). The
computation for $C_{0,i}$ is similar, but uses MinMore and does not
add the penalty $\lambda$
(lines~\ref{line:op-MinMore}--\ref{line:op-MinOfTwo-under}). The
details about how the MinLess/MinMore/MinOfTwo sub-routines process
the PiecewiseFunction objects have been described previously
\citep{Hocking-constrained-changepoint-detection}.

After computing the optimal cost functions, the decoding of optimal
parameters occurs on
lines~\ref{line:op-ArgMin}--\ref{line:op-i+1}. The last segment mean
and second to last segment end are first stored on
line~\ref{line:op-store-i} in $(U_1,T_1)$. For each other segment $i$,
the mean and previous segment end are stored on line~\ref{line:op-i+1}
in $(U_i,T_i)$. Note that there should be space to store $(U_i,T_i)$
parameters for up to $N$ segments. In practice our implementation
writes these parameters to a text output file on disk.

\paragraph{Computational complexity.} The complexity of Algorithm~\ref{algo:GFPOP} is $O(N I)$, where $I$ is
the mean number of intervals (function pieces) that are used to
represent the $C_{s, i}$ cost functions. Theoretically there are some
pathological data sets for which the algorithm computes $I=O(N)$
intervals, which results in the worst-case complexity of $O(N^2)$.
Since the number of intervals in real data is empirically
$I=O(\log N)$ (see Figure~\ref{fig:intervals}), the overall complexity
of Algorithm~\ref{algo:GFPOP} is on average $O(N \log N)$. Using
disk-based storage its complexity is $O(N\log N)$ time, $O(N\log N)$
disk, $O(\log N)$ memory. 

\paragraph{Usage in \R.} We implemented the disk-based version of Algorithm~\ref{algo:GFPOP} in
\proglang{C++} code with an interface in the \R\ package
\pkg{PeakSegDisk}. To illustrate its usage we first load a set of
genomic data,

\begin{Schunk}
\begin{Sinput}
> library(PeakSegDisk)
> data(Mono27ac, package="PeakSegDisk")
> Mono27ac$coverage
\end{Sinput}
\begin{Soutput}
      chrom chromStart chromEnd count
   1: chr11      60000   132601     0
   2: chr11     132601   132643     1
   3: chr11     132643   146765     0
   4: chr11     146765   146807     1
   5: chr11     146807   175254     0
  ---                                
6917: chr11     579752   579792     1
6918: chr11     579792   579794     2
6919: chr11     579794   579834     1
6920: chr11     579834   579980     0
6921: chr11     579980   580000     1
\end{Soutput}
\begin{Sinput}
> 
\end{Sinput}
\end{Schunk}

Note that the 4 column bedGraph format shown above must be used to
represent a data set. Futhermore a run-length encoding should be used
for data sets that have runs of the same values. Each row represents a
sequence of identical data values. For example the first row means
that the value 0 occurs on the 72601 positions in [60000,132601), the
second row means a value of 1 for the 42 positions in [132601,132643),
etc. This run-length encoding results in significant savings in disk
space and time 
\citep{Segmentor}; for
example in the data set above there are only 6921 lines used to
represent 520000 data values.

In order to handle very large data sets while using only $O(\log N)$
memory, the algorithm reads input data from a text file on disk (and
never actually stores the entire data set in memory). So before using
the algorithm we must save the data set to disk, in bedGraph format
(the four columns shown above, separated by tabs). Note that the file
name must be \verb|coverage.bedGraph|:

\begin{Schunk}
\begin{Sinput}
> data.dir <- file.path("Mono27ac", "chr11:60000-580000")
> dir.create(data.dir, showWarnings=FALSE, recursive=TRUE)
> write.table(
+   Mono27ac$coverage, file.path(data.dir, "coverage.bedGraph"),
+   col.names=FALSE, row.names=FALSE, quote=FALSE, sep="\t")
> 
\end{Sinput}
\end{Schunk}

After saving the file to disk, we can run the algorithm using the code
below: 

\begin{Schunk}
\begin{Sinput}
> ## Compute one model with penalty=10000
> fit <- PeakSegDisk::problem.PeakSegFPOP(data.dir, "10000")
> 
\end{Sinput}
\end{Schunk}

For the first argument you must give the folder name (not the
\verb|coverage.bedGraph| file name) to the problem.PeakSegFPOP
function.  Note that the second argument must be a character string
that represents a penalty value (non-negative real number, larger
penalties yield fewer peaks). The smallest value is \code{"0"} which
yields max peaks, and the largest value is \code{"Inf"} which yields
no peaks. It must be an \R\ character string (not a real number)
because that string is used to create files 
which are used to store/cache the results. If the files already exist
(and are consistent) then problem.PeakSegFPOP just reads them;
otherwise it runs the dynamic programming C++ code in order to create
those files.

The returned \code{fit} object is a named list of data.tables. The
\code{fit\$loss} component shown below is one row that contains general
information about the computed model: 

\begin{Schunk}
\begin{Sinput}
> fit$loss
\end{Sinput}
\begin{Soutput}
   penalty segments peaks  bases mean.pen.cost total.loss equality.constraints
1:   10000       15     7 520000     0.2189332   43845.26                    0
   mean.intervals max.intervals
1:       14.42581            41
\end{Soutput}
\begin{Sinput}
> 
\end{Sinput}
\end{Schunk}

Above we can see that the optimal model for $\lambda=10^4$ had $P=7$
peaks ($K=15$ segments/$K-1=14$ changepoints). The
\code{fit\$segments} component is used to visualize three of those
peaks in a subset of the data below.

\begin{Schunk}
\begin{Sinput}
> library(ggplot2)
> gg <- ggplot()+theme_bw()+
+   geom_step(aes(chromStart, count), color="grey50", data=Mono27ac$coverage)+
+   geom_segment(aes(chromStart, mean, xend=chromEnd, yend=mean),
+     color="green", size=1, data=fit$segments)+
+   coord_cartesian(xlim=c(2e5, 3e5))
> print(gg)
> 
\end{Sinput}
\end{Schunk}
\includegraphics{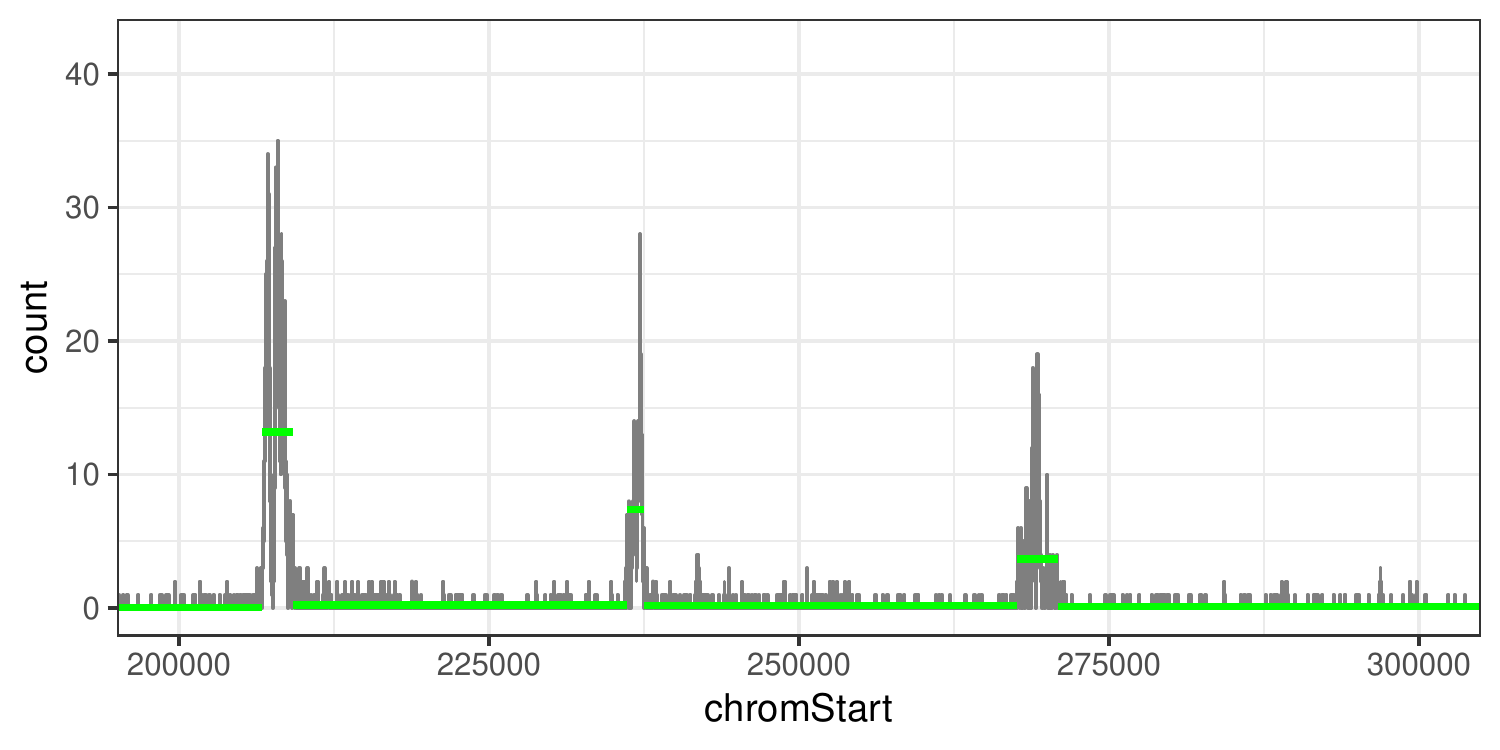}

\subsection{Sequential search algorithm for $P^*$ peaks}

Note that in GFPOP~(Algorithm~\ref{algo:GFPOP}), the user inputs a
penalty $\lambda$, and is unable to directly choose the number of
segments/peaks. In this section, we propose an algorithm that allows
the user to specify the number of peaks. The algorithm then repeatedly
calls GFPOP until it finds the most likely model with at most the
specified number of peaks.

To understand how the algorithm works, we must review the relationship
between the Optimal Partitioning and Segment Neighborhood problems
\citep{Maidstone2016}. We define the optimal loss for a given
number of peaks $P$ as
\begin{align}
  \label{eq:L_P}
  L_P=\min_{
  \substack{
  \mathbf m\in\RR^N,\ \mathbf s\in\{0, 1\}^N\\
  \mathbf c\in\{-1, 0,1\}^{N-1}\\
  }
  } &\ \ 
      \sum_{i=1}^N \ell(m_i, z_i) \\
  \text{subject to\ \ } &\ c_i = 0 \Rightarrow m_i = m_{i+1}\text{ and }s_i=s_{i+1}, \nonumber\\
    &\ c_i = 1 \Rightarrow m_i \leq m_{i+1}\text{ and }(s_i,s_{i+1})=(0,1),\nonumber\\
    &\ c_i = -1 \Rightarrow m_i \geq m_{i+1}\text{ and }(s_i,s_{i+1})=(1,0),\nonumber\\
  & \  s_1=s_N=0,\nonumber\\
& P = \sum_{i=1}^{N-1} I(c_i = 1).\label{P-constraint}
\end{align}
The problem above is the Segment Neighborhood version of the Optimal
Partitioning problem that GFPOP solves (\ref{min:op-up-down}). The
penalty $\lambda$ is absent, and the model complexity (the number of
peaks) has moved to a constraint (\ref{P-constraint}). Recall that
$F(\lambda)$ is the minimum value of the Optimal Partitioning
problem~(\ref{min:op-up-down}). It can be written in terms of the
solution to the Segment Neighborhood problem~(\ref{eq:L_P}),
\begin{equation}
  \label{eq:F(lambda)}
  F(\lambda) = \min_{
    P\in\{0,1,\dots,P_{\text{max}}\} 
  } L_P + \lambda P.
\end{equation}
The equation above makes it clear that there are only a finite number
of optimal changepoint models (from 0 to $P_{\text{max}}$
peaks). $F(\lambda)$ is a concave, non-decreasing function that can be
computed as the minimum of a finite number of affine functions
$f_P(\lambda)=L_P+\lambda P$.

\begin{figure}[t!]
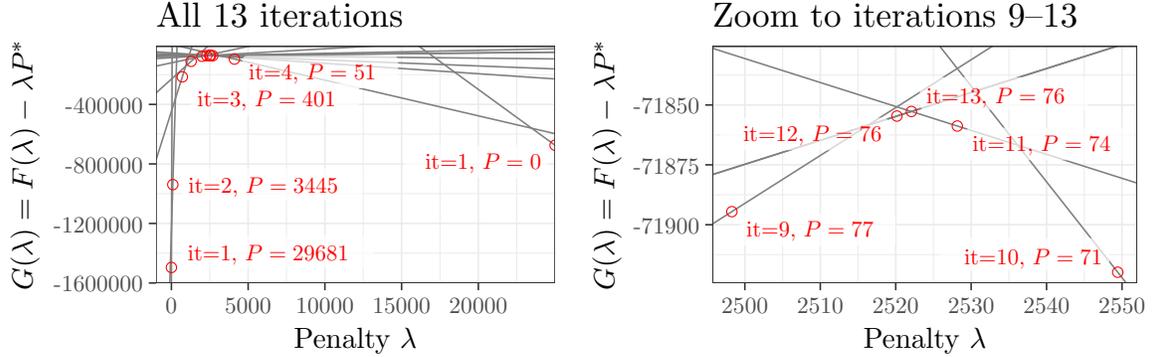

\centering
\begin{minipage}{3in}
  \input{jss-figure-evaluations-concave}
\end{minipage} 
\begin{minipage}{3in} 
  \input{jss-figure-evaluations-concave-zoom} 
\end{minipage} 
\vskip -0.5cm
\caption{\label{fig:evaluations-concave} Example of a $G(\lambda)$
  function which is maximized in order to find the most likely model
  with at most $P^*=75$ peaks. Red dots show $G(\lambda)$ values
  evaluated by the algorithm; grey lines show affine functions
  $g_P(\lambda)=L_P+(P-P^*)\lambda$ used to determine the next
  $\lambda$ value (line~\ref{line:lambda} of
  Algorithm~\ref{algo:seq-search}). \textbf{Left:} iteration 1 runs
  GFPOP with $\lambda\in\{0,\infty\}$, resulting in initial lower
  bound of $\underline p=0$ peaks and upper bound of
  $\overline p=29681$ peaks. In iteration 2 the algorithm finds the
  intersection of the upper/lower bound lines
  $g_0(\lambda)=g_{29681}(\lambda)$ at $\lambda=90.9$; running GFPOP
  with that penalty reduces the upper bound to $\underline
  p=3445$. \textbf{Right:} In the last iteration (13), we run GFPOP
  with $\lambda=2522.1$ (which is where $g_{74}$
  intersects $g_{76}$), resulting in 76 peaks when we already have
  $\overline p=76$ as an upper bound (computed in iteration 12). The
  maximum of $G$ is thus
  $G(2522.1)=g_{74}(2522.1)=g_{76}(2522.1)$; the algorithm returns the
  model with $P=74$ peaks.}
\end{figure}

We now assume the user wants to compute the optimal model with a fixed
number of peaks $P^*$. To compute that model we will maximize the function
\begin{equation}
  \label{eq:G}
  G(\lambda)=F(\lambda) - P^*\lambda = \min_{
    P\in\{0,1,\dots,P_{\text{max}}\}
    } \underbrace{L_P + \lambda(P-P^*)}_{g_P(\lambda)}.
\end{equation}
From the equation above it is clear that $G(\lambda)$ is a concave
function that can be computed as the minimum of a finite number of
affine functions $g_P(\lambda)=L_P+\lambda(P-P^*)$. 
For an example $G$
function see Figure~\ref{fig:evaluations-concave}.

\paragraph{Discussion of pseudocode.} Algorithm~\ref{algo:seq-search}
summarizes the sequential search. The main idea of the sequential
search algorithm is to keep track of a lower bound $\underline p<P^*$
and upper bound $\overline p>P^*$ on the number of peaks computed thus
far.  The algorithm starts with $\lambda=0$, $\overline p=P_{\text{max}}$
(line~\ref{line:lambda0}) and $\lambda=\infty,\underline p=0$
(line~\ref{line:lambdaInf}).  At each iteration of the algorithm, we
find the intersection of the affine functions
$g_{\underline p}(\lambda)=g_{\overline p}(\lambda)$, which leads to a
new candidate penalty
$\lambda_{\text{new}}=(L_{\overline p}-L_{\underline
  p})/(\underline p-\overline p)$ (line~\ref{line:lambda}).
As previously described
\citep{crops}, there are two possibilities for the solution to the
Optimal Partitioning problem:
\begin{itemize}
\item GFPOP($\lambda_{\text{new}}$) yields
  $\underline p$ or $\overline p$ peaks (line~\ref{line:returnFewer}). In that case
  $\max_\lambda G(\lambda)=G(\lambda_{\text{new}})=g_{\underline p}(\lambda_{\text{new}})=g_{\overline p}(\lambda_{\text{new}})$ and there
  is no Optimal Partitioning model with $P^*$ peaks. We terminate the
  algorithm by returning the model with $\underline p$ peaks.
\item GFPOP($\lambda_{\text{new}}$) yields a new model
  with $p_{\text{new}}$ peaks. If $p_{\text{new}}=P^*$ then
  $\max_\lambda G(\lambda)=L_{P^*}$ and we return this
  model (line~\ref{line:returnNew}). Otherwise it must be true that
  $\underline p < p_{\text{new}} < \overline p$. If $\underline p<p_{\text{new}} < P^*$
  then we use $p_{\text{new}}$ for a new lower bound $\underline p$ (line~\ref{line:new-under}); otherwise
  we use it for a new upper bound $\overline p$ (line~\ref{line:new-over}).
\end{itemize}

\begin{algorithm}
    \begin{algorithmic}[1]
  \STATE Input: data $\mathbf z\in\RR^N$, target peaks $P^*$.
  \STATE $\overline L,\overline p \gets \text{GFPOP}(\mathbf z, \lambda=0)$ // initialize upper bound to max peak model
  \label{line:lambda0}
  \STATE $\underline L,\underline p \gets \text{GFPOP}(\mathbf z, \lambda=\infty)$ // initialize lower bound to 0 peak model
  \label{line:lambdaInf}
  \STATE While $P^*\not\in \{\underline p,\, \overline p\}$:
  \label{line:while}
  \begin{ALC@g}
    \STATE $\lambda_{\text{new}}=(\overline L-\underline L)/(\underline p-\overline p)$
    \label{line:lambda}
    \STATE $L_{\text{new}},p_{\text{new}}\gets\text{GFPOP}(\mathbf z, \lambda_{\text{new}})$
    \STATE If $p_{\text{new}}\in\{\underline p, \overline p\}$: return model with $\underline p$ peaks.
    \label{line:returnFewer}
    \STATE If $p_{\text{new}}=P^*$: return model with $p_{\text{new}}$ peaks.
    \label{line:returnNew}
    \STATE If $p_{\text{new}} < P^*$: $\underline L,\underline p\gets L_{\text{new}},p_{\text{new}}$ // new lower bound
    \label{line:new-under}
    \STATE Else: $\overline L,\overline p\gets L_{\text{new}},p_{\text{new}}$ // new upper bound
    \label{line:new-over}
  \end{ALC@g}
  \end{algorithmic}
\caption{\label{algo:seq-search}Sequential search for $P^*$ peaks using GFPOP.}
\end{algorithm}

\paragraph{Computational complexity.} The space complexity is the same as
GFPOP: $O(N\log N)$ disk and $O(\log N)$ memory. Its time complexity
is linear in the number of iterations of the while loop
(line~\ref{line:while}). Empirically we see $O(\log P^*)$ iterations
(Section~\ref{sec:op-faster-than-sn}), which implies an overall time
complexity of $O(N\log(N)\log(P^*))$.

\paragraph{Usage in \R.} The \R\ code below computes the optimal model with 17 peaks:

\begin{Schunk}
\begin{Sinput}
> ## Compute the optimal model with 17 peaks.
> fit <- PeakSegDisk::problem.sequentialSearch(data.dir, 17L)
> 
\end{Sinput}
\end{Schunk}

If you want to see how many iterations/penalties the algorithm
required in order to compute the optimal model with 17 peaks, you can
look at the \verb|fit$others| component:

\begin{Schunk}
\begin{Sinput}
> fit$others[, list(iteration, under, over, penalty, peaks, total.loss)]
\end{Sinput}
\begin{Soutput}
   iteration under over   penalty peaks  total.loss
1:         1    NA   NA    0.0000  3199 -130227.291
2:         1    NA   NA       Inf     0  375197.873
3:         2     0 3199  157.9947   224  -62199.931
4:         3     0  224 1952.6688    17    2640.128
\end{Soutput}
\begin{Sinput}
> 
\end{Sinput}
\end{Schunk}

The output above shows that the algorithm only used three iterations
to compute the optimal model with 17 peaks. The \code{under} and
\code{over} columns show the current values of $\underline p$ and
$\overline p$, respectively. The \code{peaks} and \code{total.loss}
are $p_{\text{new}},L_{\text{new}}$ from the model that resulted from running GFPOP
with $\lambda=\text{\code{penalty}}$. Note that iteration 1 evaluates both
extreme penalties $\lambda\in\{0,\infty\}$ in parallel (and
$\lambda=\infty$ is the trivial model with 0 peaks that can be
computed without dynamic programming), so these two models are
considered a single iteration.

\section{Results on genomic data} \label{sec:results}

In this section we discuss several applications of our algorithms in
some typical genomic data sets. We downloaded the \verb|chipseq| data set from
the UCI machine learning repository \citep{uci-ml-repo}. We considered
4951 data sets ranging in size from $N=10^3$ to $N=10^7$ data points
to segment (lines in the bedGraph file). 

\begin{figure}[t!]
\centering
\begin{minipage}{3.5in}
  \includegraphics[width=\textwidth]{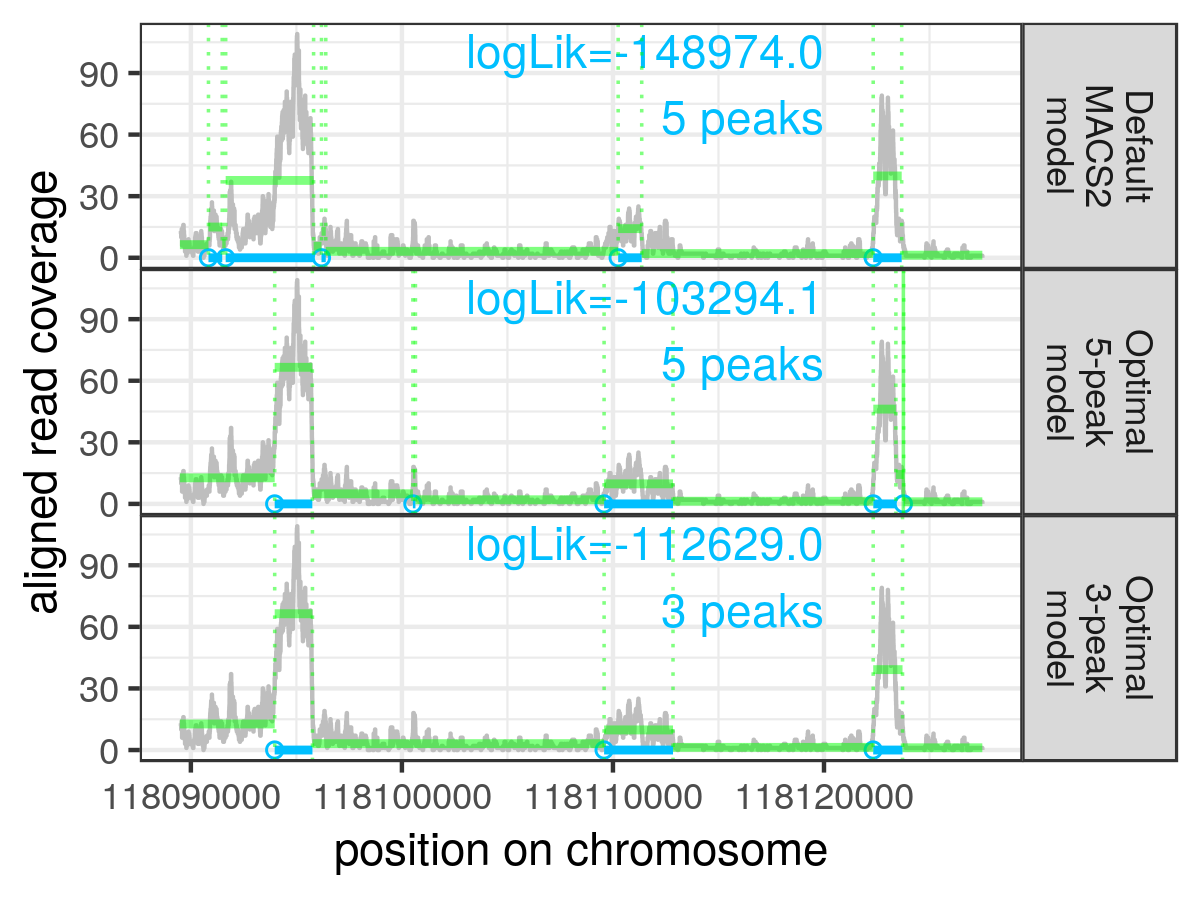}
\end{minipage}
\begin{minipage}{2.3in}
  \includegraphics[width=\textwidth]{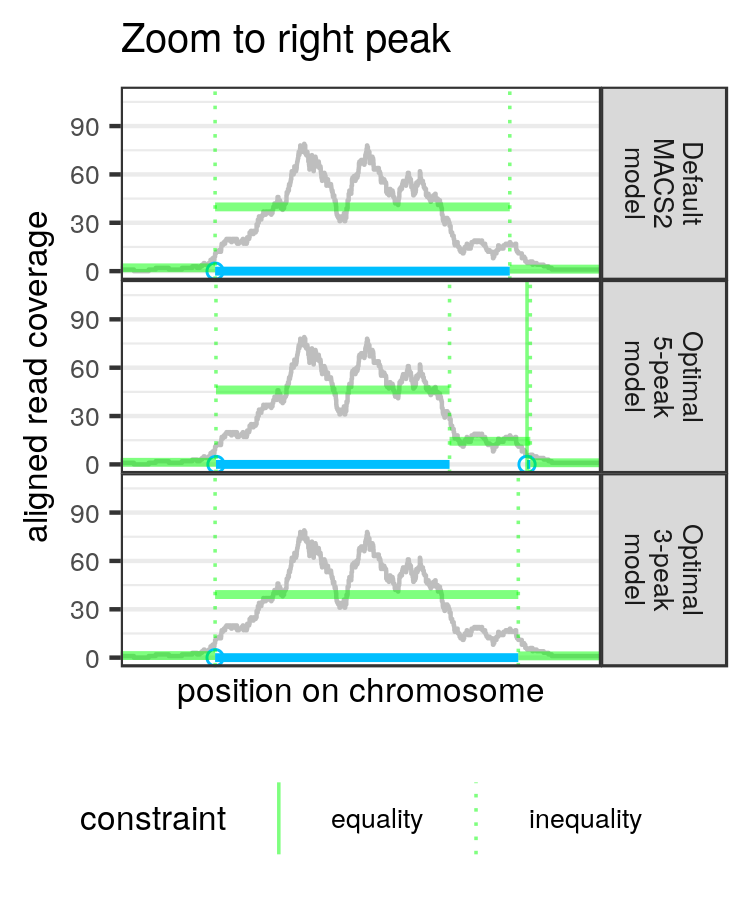}
\end{minipage}
\caption{\label{fig:three-peaks} One ChIP-seq data set with three peak
  models. (green horizontal segment means; green dotted vertical lines
  for changepoints; blue bars for peaks; blue dots for peak starts)
  \textbf{Top:} the MACS2 algorithm (a heuristic from the
  bioinformatics literature) computed a sub-optimal model with five
  peaks for these data. \textbf{Middle:} the most likely model with
  five peaks contains one equality constraint between segment means
  (see zoomed figure on the right), which suggests that there are less
  than five easily interpretable peaks. \textbf{Bottom:} the most
  likely model with three peaks is also more likely than the MACS2
  model.}
\end{figure}
 
\subsection{Application: Computing the maximum likelihood model with a
  given number of peaks}

In this section we show that our algorithms can be used to compute the
most likely model for a given number of peaks. A subset of one data
set is shown in Figure~\ref{fig:three-peaks}, along with three
segmentation/peak models. In the top panel, we show the peak model
that results from running MACS2, a heuristic algorithm from the
bioinformatics literature \citep{MACS}. It detects five peaks, so we
ran Algorithm~\ref{algo:seq-search} with $P^*=5$ on these data in
order to compute the most likely model with at most 5 peaks (shown in
middle panel). It is clear that the optimal 5 peak model is a better
fit in terms of likelihood (as expected); it is also a better fit
visually, especially for the peak on the left. Furthermore the optimal
5 peak model actually has one equality constraint between adjacent
segment means, suggesting that there are less than five easily
interpretable peaks. Therefore we also computed the optimal 3 peak
model (bottom panel), which also has a higher log-likelihood than the
5 peak MACS2 model. Overall it is clear that our algorithms can be
used to compute models which are both more likely and simpler (with
fewer peaks) than heuristics such as MACS2.


\subsection{GFPOP is empirically log-linear}

\begin{figure}[t!]
\centering
\begin{minipage}{3in}
  \includegraphics{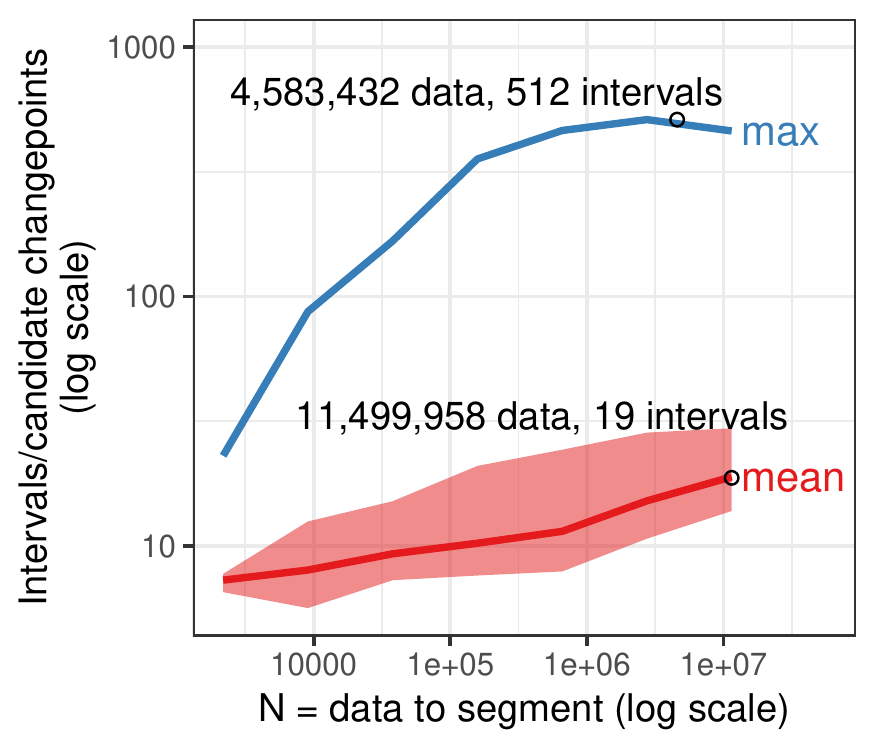}
\end{minipage} 
\begin{minipage}{3in}
  \includegraphics{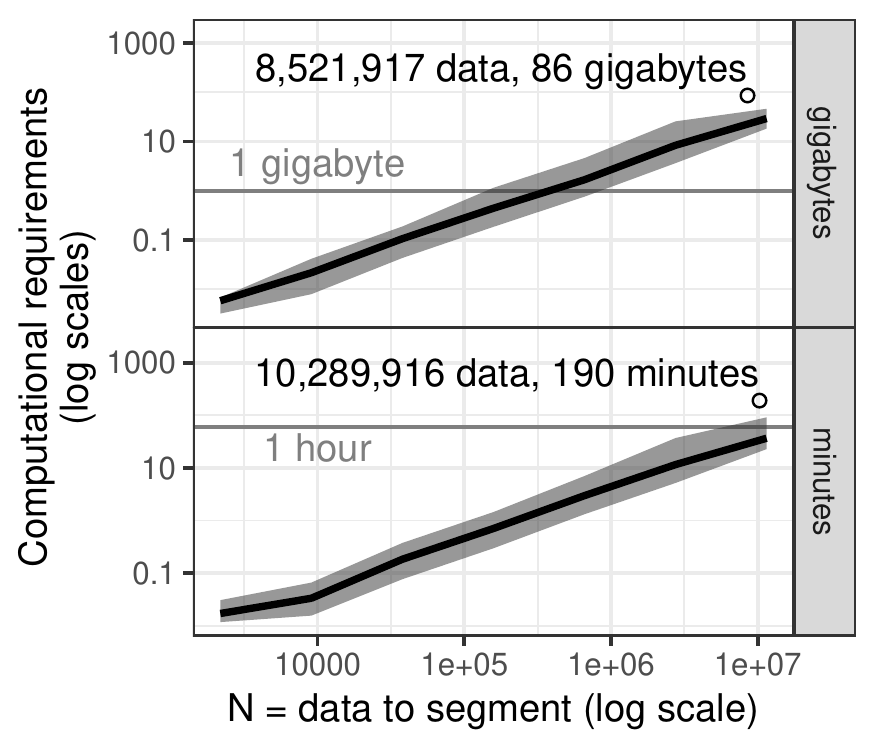}
\end{minipage}
\caption{\label{fig:intervals} In our empirical
  tests, the computational requirements of the GFPOP algorithm
  were log-linear $O(N \log N)$ in the number of data points $N$ to
  segment. \textbf{Left:} we analyzed the number of intervals $I$
  (candidate changepoints) stored in the $C_t(\mu)$ cost functions,
  because the total time/space complexity is $O(NI)$. We observed
  empirically that the mean number of intervals $I=O(\log N)$ (red
  curve). Even the maximum number of intervals (blue curve) is much
  less than $N$. \textbf{Right:} storage on disk (top panel) and
  computation time (bottom panel) are empirically $O(N \log N)$. Error
  bands show median and 5\%/95\% quantiles over several data sets of a
  given size $N$; black dots and text show computational requirements
  for the most extreme data sets.}
\end{figure}

To measure the empirical time complexity of GFPOP
(Algorithm~\ref{algo:GFPOP}), we ran it on all 4951 genomic data sets,
with a grid of penalty values $\lambda\in(\log N, N)$ for each data
set of size $N$. The overall theoretical time/space complexity is
$O(NI)$, where $I$ is the number of intervals (candidate changepoints)
stored in the $C_{s,t}$ optimal cost functions. During each run we
therefore recorded the mean and max number of intervals over all
$s,t$. We observed that the empirical mean/max number of intervals
increases logarithmically with data set size, $I=O(\log N)$
(Figure~\ref{fig:intervals}, left). Remarkably, for the largest data
set ($N=11,499,958$) the algorithm only computed a mean of $I=19$
intervals. The most intervals computed to represent any single
$C_{s,t}$ function was 512 intervals for one data set with
$N=4,583,432$. 

Since empirically $I=O(\log N)$ in these genomic data sets, we
expected an overall time/space complexity of $O(N\log N)$. The
empirical measurements of time and space requirements are consistent
with this expectation (Figure~\ref{fig:intervals}, right). For the
largest data sets ($N=10^7$), the algorithm takes only about 80
gigabytes of storage and 1 hour of computation time. Overall these
results suggest that GFPOP can be used to compute optimal peak models
for genomic data in $O(N\log N)$ space/time.
 
\subsection{Disk storage is slower than memory by a constant
  factor}

\begin{figure}[t!]
\centering
\begin{minipage}{3in}
  \includegraphics{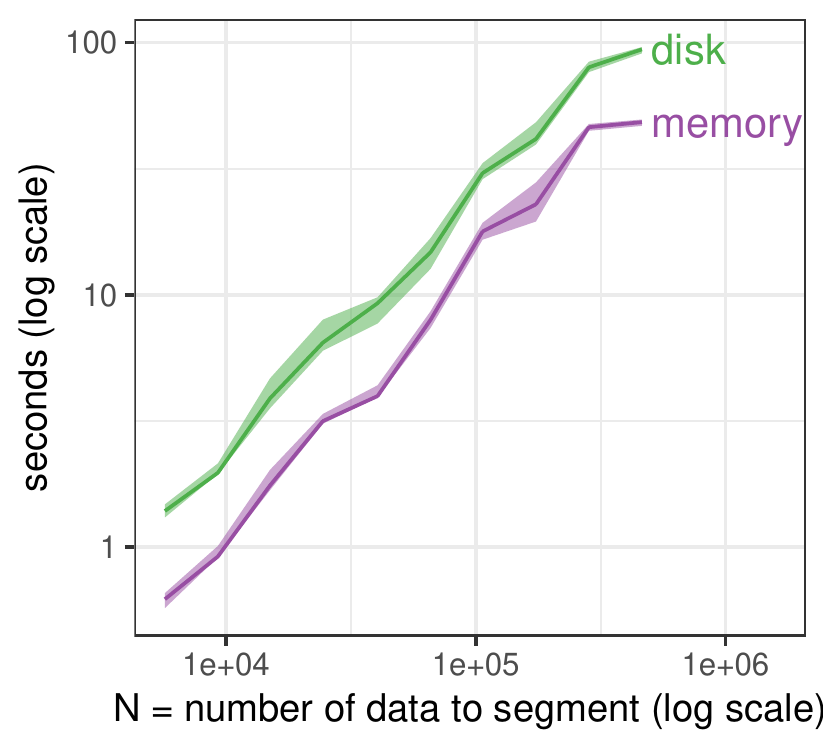}
\end{minipage} 
\begin{minipage}{3in}
  \includegraphics{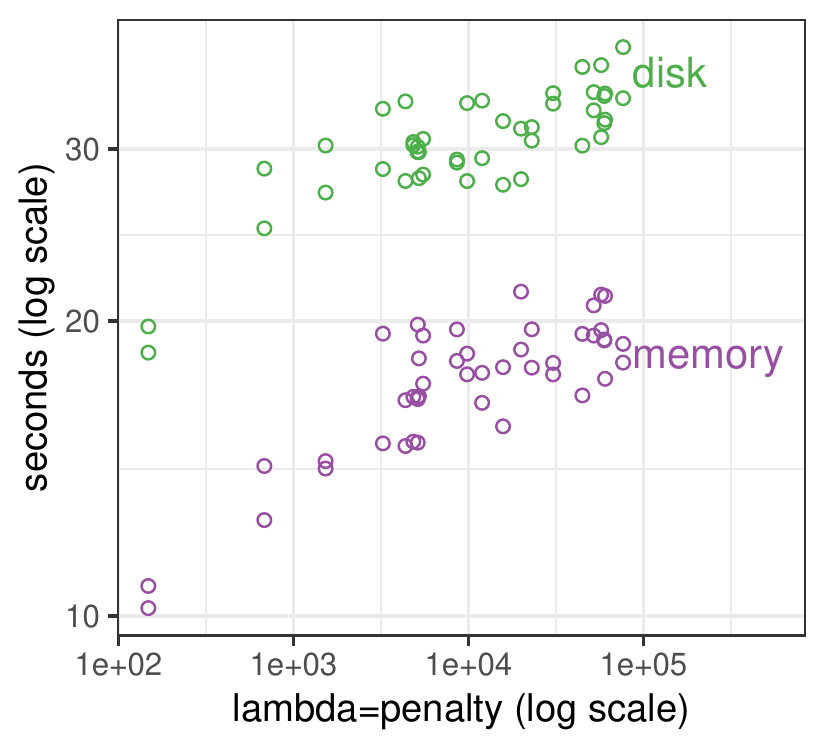}
\end{minipage}
\caption{\label{fig:disk-memory-compare-speed} The disk-based storage
  method is only a constant factor slower than the memory-based
  method. We benchmarked both methods on several small data sets
  ($N\leq 462,890$) for which optimal models could be computed using
  1GB of storage. \textbf{Left:} computation time is empirically
  $O(N\log N)$ for both storage methods, but the disk-based method is
  slower by a constant factor. Median line and quartile band computed
  over several penalty values for a given data set. \textbf{Right:}
  fixing one data set data set with $N=106,569$, the computation time
  increases with penalty value $\lambda$ for both storage methods.}
\end{figure}

In the previous section, we discussed how tens of gigabytes of
storage are required to run GFPOP when $N=10^7$. Since typical
computers may not have enough memory, our implementation uses
disk-based storage. We compared our disk-based implementation to
another memory-based implementation, in terms of computation time on
small data sets for which GFPOP uses $<1$GB of storage. We observed
that disk storage is slower than memory storage by a constant factor
(1.7--2.3$\times$, Figure~\ref{fig:disk-memory-compare-speed}), which was expected. 

\subsection{Sequential search is faster than Segment Neighborhood}
\label{sec:op-faster-than-sn}
In this section we compare the number of $O(N\log N)$ dynamic
programming iterations required for the proposed sequential search
(Algorithm~\ref{algo:seq-search}) and the previous Generalized Pruned
Dynamic Programming Algorithm (GPDPA) of
\citet{Hocking-constrained-changepoint-detection}. Both algorithms
compute the solution to the Segment Neighborhood problem (optimal
model with at most $P$ peaks). The GPDPA requires exactly $2P$ iterations of
dynamic programming, each of which is an $O(N\log N)$ operation. In
contrast, the proposed sequential search
(Algorithm~\ref{algo:seq-search}) needs to solve for a sequence of
penalties, each of which is done via GFPOP in $O(N\log N)$ time. 

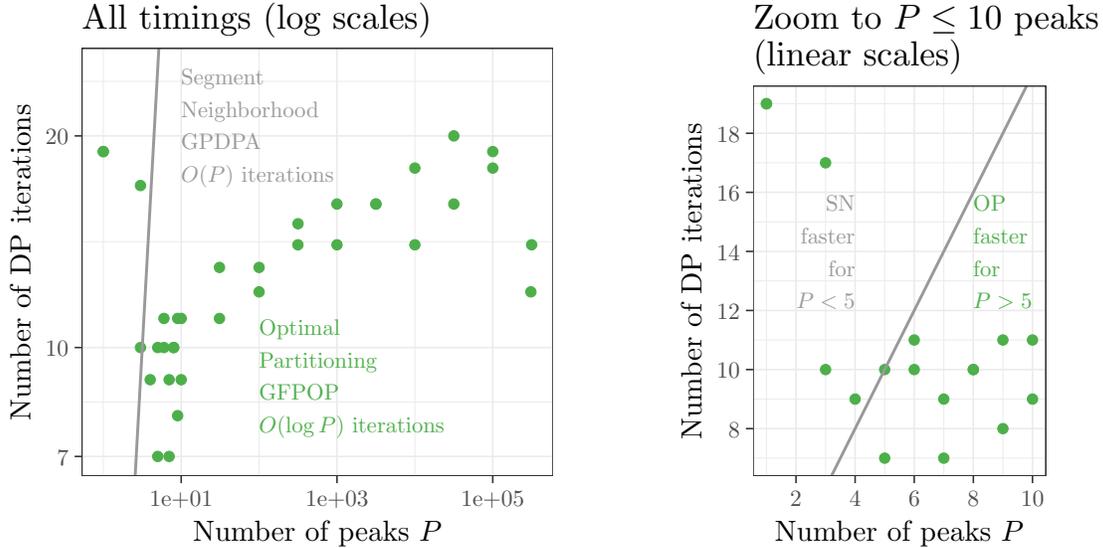
\begin{figure}[t!]
\centering
\begin{minipage}{3in}
\begin{tikzpicture}[x=1pt,y=1pt]
\definecolor{fillColor}{RGB}{255,255,255}
\path[use as bounding box,fill=fillColor,fill opacity=0.00] (0,0) rectangle (216.81,216.81);
\begin{scope}
\path[clip] (  0.00,  0.00) rectangle (216.81,216.81);
\definecolor{drawColor}{RGB}{255,255,255}
\definecolor{fillColor}{RGB}{255,255,255}

\path[draw=drawColor,line width= 0.6pt,line join=round,line cap=round,fill=fillColor] (  0.00,  0.00) rectangle (216.81,216.81);
\end{scope}
\begin{scope}
\path[clip] ( 33.09, 32.08) rectangle (211.31,193.93);
\definecolor{fillColor}{RGB}{255,255,255}

\path[fill=fillColor] ( 33.09, 32.08) rectangle (211.31,193.93);
\definecolor{drawColor}{gray}{0.92}

\path[draw=drawColor,line width= 0.3pt,line join=round] ( 33.09, 60.05) --
	(211.31, 60.05);

\path[draw=drawColor,line width= 0.3pt,line join=round] ( 33.09,120.72) --
	(211.31,120.72);

\path[draw=drawColor,line width= 0.3pt,line join=round] ( 33.09,181.39) --
	(211.31,181.39);

\path[draw=drawColor,line width= 0.3pt,line join=round] ( 41.19, 32.08) --
	( 41.19,193.93);

\path[draw=drawColor,line width= 0.3pt,line join=round] (100.11, 32.08) --
	(100.11,193.93);

\path[draw=drawColor,line width= 0.3pt,line join=round] (159.04, 32.08) --
	(159.04,193.93);

\path[draw=drawColor,line width= 0.6pt,line join=round] ( 33.09, 39.44) --
	(211.31, 39.44);

\path[draw=drawColor,line width= 0.6pt,line join=round] ( 33.09, 80.67) --
	(211.31, 80.67);

\path[draw=drawColor,line width= 0.6pt,line join=round] ( 33.09,160.78) --
	(211.31,160.78);

\path[draw=drawColor,line width= 0.6pt,line join=round] ( 70.65, 32.08) --
	( 70.65,193.93);

\path[draw=drawColor,line width= 0.6pt,line join=round] (129.58, 32.08) --
	(129.58,193.93);

\path[draw=drawColor,line width= 0.6pt,line join=round] (188.50, 32.08) --
	(188.50,193.93);
\definecolor{drawColor}{RGB}{77,175,74}
\definecolor{fillColor}{RGB}{77,175,74}

\path[draw=drawColor,line width= 0.4pt,line join=round,line cap=round,fill=fillColor] ( 41.19,154.85) circle (  1.96);

\path[draw=drawColor,line width= 0.4pt,line join=round,line cap=round,fill=fillColor] ( 55.25, 80.67) circle (  1.96);

\path[draw=drawColor,line width= 0.4pt,line join=round,line cap=round,fill=fillColor] ( 58.93, 68.49) circle (  1.96);

\path[draw=drawColor,line width= 0.4pt,line join=round,line cap=round,fill=fillColor] ( 61.78, 80.67) circle (  1.96);

\path[draw=drawColor,line width= 0.4pt,line join=round,line cap=round,fill=fillColor] ( 64.12, 91.68) circle (  1.96);

\path[draw=drawColor,line width= 0.4pt,line join=round,line cap=round,fill=fillColor] ( 66.09, 39.44) circle (  1.96);

\path[draw=drawColor,line width= 0.4pt,line join=round,line cap=round,fill=fillColor] ( 67.80, 80.67) circle (  1.96);

\path[draw=drawColor,line width= 0.4pt,line join=round,line cap=round,fill=fillColor] ( 69.31, 91.68) circle (  1.96);

\path[draw=drawColor,line width= 0.4pt,line join=round,line cap=round,fill=fillColor] ( 70.65, 68.49) circle (  1.96);

\path[draw=drawColor,line width= 0.4pt,line join=round,line cap=round,fill=fillColor] ( 85.13,110.99) circle (  1.96);

\path[draw=drawColor,line width= 0.4pt,line join=round,line cap=round,fill=fillColor] (100.11,101.74) circle (  1.96);

\path[draw=drawColor,line width= 0.4pt,line join=round,line cap=round,fill=fillColor] (114.84,127.53) circle (  1.96);

\path[draw=drawColor,line width= 0.4pt,line join=round,line cap=round,fill=fillColor] (129.56,134.99) circle (  1.96);

\path[draw=drawColor,line width= 0.4pt,line join=round,line cap=round,fill=fillColor] (144.31,134.99) circle (  1.96);

\path[draw=drawColor,line width= 0.4pt,line join=round,line cap=round,fill=fillColor] (159.04,148.60) circle (  1.96);

\path[draw=drawColor,line width= 0.4pt,line join=round,line cap=round,fill=fillColor] (173.77,160.78) circle (  1.96);

\path[draw=drawColor,line width= 0.4pt,line join=round,line cap=round,fill=fillColor] (188.50,148.60) circle (  1.96);

\path[draw=drawColor,line width= 0.4pt,line join=round,line cap=round,fill=fillColor] (203.21,119.56) circle (  1.96);

\path[draw=drawColor,line width= 0.4pt,line join=round,line cap=round,fill=fillColor] ( 41.19,154.85) circle (  1.96);

\path[draw=drawColor,line width= 0.4pt,line join=round,line cap=round,fill=fillColor] ( 55.25,142.00) circle (  1.96);

\path[draw=drawColor,line width= 0.4pt,line join=round,line cap=round,fill=fillColor] ( 61.78, 39.44) circle (  1.96);

\path[draw=drawColor,line width= 0.4pt,line join=round,line cap=round,fill=fillColor] ( 64.12, 80.67) circle (  1.96);

\path[draw=drawColor,line width= 0.4pt,line join=round,line cap=round,fill=fillColor] ( 66.09, 68.49) circle (  1.96);

\path[draw=drawColor,line width= 0.4pt,line join=round,line cap=round,fill=fillColor] ( 67.80, 80.67) circle (  1.96);

\path[draw=drawColor,line width= 0.4pt,line join=round,line cap=round,fill=fillColor] ( 69.31, 54.87) circle (  1.96);

\path[draw=drawColor,line width= 0.4pt,line join=round,line cap=round,fill=fillColor] ( 70.65, 91.68) circle (  1.96);

\path[draw=drawColor,line width= 0.4pt,line join=round,line cap=round,fill=fillColor] ( 85.13, 91.68) circle (  1.96);

\path[draw=drawColor,line width= 0.4pt,line join=round,line cap=round,fill=fillColor] (100.11,110.99) circle (  1.96);

\path[draw=drawColor,line width= 0.4pt,line join=round,line cap=round,fill=fillColor] (114.80,119.56) circle (  1.96);

\path[draw=drawColor,line width= 0.4pt,line join=round,line cap=round,fill=fillColor] (129.58,119.56) circle (  1.96);

\path[draw=drawColor,line width= 0.4pt,line join=round,line cap=round,fill=fillColor] (144.31,134.99) circle (  1.96);

\path[draw=drawColor,line width= 0.4pt,line join=round,line cap=round,fill=fillColor] (159.04,119.56) circle (  1.96);

\path[draw=drawColor,line width= 0.4pt,line join=round,line cap=round,fill=fillColor] (173.77,134.99) circle (  1.96);

\path[draw=drawColor,line width= 0.4pt,line join=round,line cap=round,fill=fillColor] (188.50,154.85) circle (  1.96);

\path[draw=drawColor,line width= 0.4pt,line join=round,line cap=round,fill=fillColor] (202.92,101.74) circle (  1.96);
\definecolor{drawColor}{gray}{0.60}

\path[draw=drawColor,line width= 1.1pt,line join=round] ( 51.46,  0.00) -- ( 63.46,216.81);

\node[text=drawColor,anchor=base west,inner sep=0pt, outer sep=0pt, scale=  0.85] at ( 70.65,180.26) {Segment};

\node[text=drawColor,anchor=base west,inner sep=0pt, outer sep=0pt, scale=  0.85] at ( 70.65,167.97) {Neighborhood};

\node[text=drawColor,anchor=base west,inner sep=0pt, outer sep=0pt, scale=  0.85] at ( 70.65,155.68) {GPDPA};

\node[text=drawColor,anchor=base west,inner sep=0pt, outer sep=0pt, scale=  0.85] at ( 70.65,143.39) {$O(P)$ iterations};
\definecolor{drawColor}{RGB}{77,175,74}

\node[text=drawColor,anchor=base west,inner sep=0pt, outer sep=0pt, scale=  0.85] at (100.11, 85.37) {Optimal};

\node[text=drawColor,anchor=base west,inner sep=0pt, outer sep=0pt, scale=  0.85] at (100.11, 73.08) {Partitioning};

\node[text=drawColor,anchor=base west,inner sep=0pt, outer sep=0pt, scale=  0.85] at (100.11, 60.79) {GFPOP};

\node[text=drawColor,anchor=base west,inner sep=0pt, outer sep=0pt, scale=  0.85] at (100.11, 48.50) {$O(\log P)$ iterations};
\definecolor{drawColor}{gray}{0.20}

\path[draw=drawColor,line width= 0.6pt,line join=round,line cap=round] ( 33.09, 32.08) rectangle (211.31,193.93);
\end{scope}
\begin{scope}
\path[clip] (  0.00,  0.00) rectangle (216.81,216.81);
\definecolor{drawColor}{gray}{0.30}

\node[text=drawColor,anchor=base east,inner sep=0pt, outer sep=0pt, scale=  0.88] at ( 28.14, 36.19) {7};

\node[text=drawColor,anchor=base east,inner sep=0pt, outer sep=0pt, scale=  0.88] at ( 28.14, 77.41) {10};

\node[text=drawColor,anchor=base east,inner sep=0pt, outer sep=0pt, scale=  0.88] at ( 28.14,157.53) {20};
\end{scope}
\begin{scope}
\path[clip] (  0.00,  0.00) rectangle (216.81,216.81);
\definecolor{drawColor}{gray}{0.20}

\path[draw=drawColor,line width= 0.6pt,line join=round] ( 30.34, 39.44) --
	( 33.09, 39.44);

\path[draw=drawColor,line width= 0.6pt,line join=round] ( 30.34, 80.67) --
	( 33.09, 80.67);

\path[draw=drawColor,line width= 0.6pt,line join=round] ( 30.34,160.78) --
	( 33.09,160.78);
\end{scope}
\begin{scope}
\path[clip] (  0.00,  0.00) rectangle (216.81,216.81);
\definecolor{drawColor}{gray}{0.20}

\path[draw=drawColor,line width= 0.6pt,line join=round] ( 70.65, 29.33) --
	( 70.65, 32.08);

\path[draw=drawColor,line width= 0.6pt,line join=round] (129.58, 29.33) --
	(129.58, 32.08);

\path[draw=drawColor,line width= 0.6pt,line join=round] (188.50, 29.33) --
	(188.50, 32.08);
\end{scope}
\begin{scope}
\path[clip] (  0.00,  0.00) rectangle (216.81,216.81);
\definecolor{drawColor}{gray}{0.30}

\node[text=drawColor,anchor=base,inner sep=0pt, outer sep=0pt, scale=  0.88] at ( 70.65, 20.63) {1e+01};

\node[text=drawColor,anchor=base,inner sep=0pt, outer sep=0pt, scale=  0.88] at (129.58, 20.63) {1e+03};

\node[text=drawColor,anchor=base,inner sep=0pt, outer sep=0pt, scale=  0.88] at (188.50, 20.63) {1e+05};
\end{scope}
\begin{scope}
\path[clip] (  0.00,  0.00) rectangle (216.81,216.81);
\definecolor{drawColor}{RGB}{0,0,0}

\node[text=drawColor,anchor=base,inner sep=0pt, outer sep=0pt, scale=  1.10] at (122.20,  7.62) {Number of peaks $P$};
\end{scope}
\begin{scope}
\path[clip] (  0.00,  0.00) rectangle (216.81,216.81);
\definecolor{drawColor}{RGB}{0,0,0}

\node[text=drawColor,rotate= 90.00,anchor=base,inner sep=0pt, outer sep=0pt, scale=  1.10] at ( 13.63,113.01) {Number of DP iterations};
\end{scope}
\begin{scope}
\path[clip] (  0.00,  0.00) rectangle (216.81,216.81);
\definecolor{drawColor}{RGB}{0,0,0}

\node[text=drawColor,anchor=base west,inner sep=0pt, outer sep=0pt, scale=  1.32] at ( 33.09,201.55) {All timings (log scales)};
\end{scope}
\end{tikzpicture}
\end{minipage}
\begin{minipage}{3in} 
\begin{tikzpicture}[x=1pt,y=1pt]
\definecolor{fillColor}{RGB}{255,255,255}
\path[use as bounding box,fill=fillColor,fill opacity=0.00] (0,0) rectangle (216.81,216.81);
\begin{scope}
\path[clip] ( 33.76,  0.00) rectangle (183.05,216.81);
\definecolor{drawColor}{RGB}{255,255,255}
\definecolor{fillColor}{RGB}{255,255,255}

\path[draw=drawColor,line width= 0.6pt,line join=round,line cap=round,fill=fillColor] ( 33.76,  0.00) rectangle (183.05,216.81);
\end{scope}
\begin{scope}
\path[clip] ( 66.85, 32.08) rectangle (177.55,179.67);
\definecolor{fillColor}{RGB}{255,255,255}

\path[fill=fillColor] ( 66.85, 32.08) rectangle (177.55,179.67);
\definecolor{drawColor}{gray}{0.92}

\path[draw=drawColor,line width= 0.3pt,line join=round] ( 66.85, 38.79) --
	(177.55, 38.79);

\path[draw=drawColor,line width= 0.3pt,line join=round] ( 66.85, 61.15) --
	(177.55, 61.15);

\path[draw=drawColor,line width= 0.3pt,line join=round] ( 66.85, 83.52) --
	(177.55, 83.52);

\path[draw=drawColor,line width= 0.3pt,line join=round] ( 66.85,105.88) --
	(177.55,105.88);

\path[draw=drawColor,line width= 0.3pt,line join=round] ( 66.85,128.24) --
	(177.55,128.24);

\path[draw=drawColor,line width= 0.3pt,line join=round] ( 66.85,150.60) --
	(177.55,150.60);

\path[draw=drawColor,line width= 0.3pt,line join=round] ( 66.85,172.96) --
	(177.55,172.96);

\path[draw=drawColor,line width= 0.3pt,line join=round] ( 71.89, 32.08) --
	( 71.89,179.67);

\path[draw=drawColor,line width= 0.3pt,line join=round] ( 94.25, 32.08) --
	( 94.25,179.67);

\path[draw=drawColor,line width= 0.3pt,line join=round] (116.61, 32.08) --
	(116.61,179.67);

\path[draw=drawColor,line width= 0.3pt,line join=round] (138.97, 32.08) --
	(138.97,179.67);

\path[draw=drawColor,line width= 0.3pt,line join=round] (161.33, 32.08) --
	(161.33,179.67);

\path[draw=drawColor,line width= 0.6pt,line join=round] ( 66.85, 49.97) --
	(177.55, 49.97);

\path[draw=drawColor,line width= 0.6pt,line join=round] ( 66.85, 72.33) --
	(177.55, 72.33);

\path[draw=drawColor,line width= 0.6pt,line join=round] ( 66.85, 94.70) --
	(177.55, 94.70);

\path[draw=drawColor,line width= 0.6pt,line join=round] ( 66.85,117.06) --
	(177.55,117.06);

\path[draw=drawColor,line width= 0.6pt,line join=round] ( 66.85,139.42) --
	(177.55,139.42);

\path[draw=drawColor,line width= 0.6pt,line join=round] ( 66.85,161.78) --
	(177.55,161.78);

\path[draw=drawColor,line width= 0.6pt,line join=round] ( 83.07, 32.08) --
	( 83.07,179.67);

\path[draw=drawColor,line width= 0.6pt,line join=round] (105.43, 32.08) --
	(105.43,179.67);

\path[draw=drawColor,line width= 0.6pt,line join=round] (127.79, 32.08) --
	(127.79,179.67);

\path[draw=drawColor,line width= 0.6pt,line join=round] (150.15, 32.08) --
	(150.15,179.67);

\path[draw=drawColor,line width= 0.6pt,line join=round] (172.52, 32.08) --
	(172.52,179.67);
\definecolor{drawColor}{RGB}{77,175,74}
\definecolor{fillColor}{RGB}{77,175,74}

\path[draw=drawColor,line width= 0.4pt,line join=round,line cap=round,fill=fillColor] ( 71.89,172.96) circle (  1.96);

\path[draw=drawColor,line width= 0.4pt,line join=round,line cap=round,fill=fillColor] ( 94.25, 72.33) circle (  1.96);

\path[draw=drawColor,line width= 0.4pt,line join=round,line cap=round,fill=fillColor] (105.43, 61.15) circle (  1.96);

\path[draw=drawColor,line width= 0.4pt,line join=round,line cap=round,fill=fillColor] (116.61, 72.33) circle (  1.96);

\path[draw=drawColor,line width= 0.4pt,line join=round,line cap=round,fill=fillColor] (127.79, 83.52) circle (  1.96);

\path[draw=drawColor,line width= 0.4pt,line join=round,line cap=round,fill=fillColor] (138.97, 38.79) circle (  1.96);

\path[draw=drawColor,line width= 0.4pt,line join=round,line cap=round,fill=fillColor] (150.15, 72.33) circle (  1.96);

\path[draw=drawColor,line width= 0.4pt,line join=round,line cap=round,fill=fillColor] (161.33, 83.52) circle (  1.96);

\path[draw=drawColor,line width= 0.4pt,line join=round,line cap=round,fill=fillColor] (172.52, 61.15) circle (  1.96);

\path[draw=drawColor,line width= 0.4pt,line join=round,line cap=round,fill=fillColor] ( 71.89,172.96) circle (  1.96);

\path[draw=drawColor,line width= 0.4pt,line join=round,line cap=round,fill=fillColor] ( 94.25,150.60) circle (  1.96);

\path[draw=drawColor,line width= 0.4pt,line join=round,line cap=round,fill=fillColor] (116.61, 38.79) circle (  1.96);

\path[draw=drawColor,line width= 0.4pt,line join=round,line cap=round,fill=fillColor] (127.79, 72.33) circle (  1.96);

\path[draw=drawColor,line width= 0.4pt,line join=round,line cap=round,fill=fillColor] (138.97, 61.15) circle (  1.96);

\path[draw=drawColor,line width= 0.4pt,line join=round,line cap=round,fill=fillColor] (150.15, 72.33) circle (  1.96);

\path[draw=drawColor,line width= 0.4pt,line join=round,line cap=round,fill=fillColor] (161.33, 49.97) circle (  1.96);

\path[draw=drawColor,line width= 0.4pt,line join=round,line cap=round,fill=fillColor] (172.52, 83.52) circle (  1.96);
\definecolor{drawColor}{gray}{0.60}

\path[draw=drawColor,line width= 1.1pt,line join=round] ( 80.44,  0.00) -- (177.55,194.21);

\node[text=drawColor,anchor=base east,inner sep=0pt, outer sep=0pt, scale=  0.85] at (105.43,132.34) {SN};

\node[text=drawColor,anchor=base east,inner sep=0pt, outer sep=0pt, scale=  0.85] at (105.43,120.05) {faster};

\node[text=drawColor,anchor=base east,inner sep=0pt, outer sep=0pt, scale=  0.85] at (105.43,107.76) {for};

\node[text=drawColor,anchor=base east,inner sep=0pt, outer sep=0pt, scale=  0.85] at (105.43, 95.47) {$P<5$};
\definecolor{drawColor}{RGB}{77,175,74}

\node[text=drawColor,anchor=base west,inner sep=0pt, outer sep=0pt, scale=  0.85] at (150.15,132.34) {OP};

\node[text=drawColor,anchor=base west,inner sep=0pt, outer sep=0pt, scale=  0.85] at (150.15,120.05) {faster};

\node[text=drawColor,anchor=base west,inner sep=0pt, outer sep=0pt, scale=  0.85] at (150.15,107.76) {for};

\node[text=drawColor,anchor=base west,inner sep=0pt, outer sep=0pt, scale=  0.85] at (150.15, 95.47) {$P>5$};
\definecolor{drawColor}{gray}{0.20}

\path[draw=drawColor,line width= 0.6pt,line join=round,line cap=round] ( 66.85, 32.08) rectangle (177.55,179.67);
\end{scope}
\begin{scope}
\path[clip] (  0.00,  0.00) rectangle (216.81,216.81);
\definecolor{drawColor}{gray}{0.30}

\node[text=drawColor,anchor=base east,inner sep=0pt, outer sep=0pt, scale=  0.88] at ( 61.90, 46.72) {8};

\node[text=drawColor,anchor=base east,inner sep=0pt, outer sep=0pt, scale=  0.88] at ( 61.90, 69.08) {10};

\node[text=drawColor,anchor=base east,inner sep=0pt, outer sep=0pt, scale=  0.88] at ( 61.90, 91.45) {12};

\node[text=drawColor,anchor=base east,inner sep=0pt, outer sep=0pt, scale=  0.88] at ( 61.90,113.81) {14};

\node[text=drawColor,anchor=base east,inner sep=0pt, outer sep=0pt, scale=  0.88] at ( 61.90,136.17) {16};

\node[text=drawColor,anchor=base east,inner sep=0pt, outer sep=0pt, scale=  0.88] at ( 61.90,158.53) {18};
\end{scope}
\begin{scope}
\path[clip] (  0.00,  0.00) rectangle (216.81,216.81);
\definecolor{drawColor}{gray}{0.20}

\path[draw=drawColor,line width= 0.6pt,line join=round] ( 64.10, 49.97) --
	( 66.85, 49.97);

\path[draw=drawColor,line width= 0.6pt,line join=round] ( 64.10, 72.33) --
	( 66.85, 72.33);

\path[draw=drawColor,line width= 0.6pt,line join=round] ( 64.10, 94.70) --
	( 66.85, 94.70);

\path[draw=drawColor,line width= 0.6pt,line join=round] ( 64.10,117.06) --
	( 66.85,117.06);

\path[draw=drawColor,line width= 0.6pt,line join=round] ( 64.10,139.42) --
	( 66.85,139.42);

\path[draw=drawColor,line width= 0.6pt,line join=round] ( 64.10,161.78) --
	( 66.85,161.78);
\end{scope}
\begin{scope}
\path[clip] (  0.00,  0.00) rectangle (216.81,216.81);
\definecolor{drawColor}{gray}{0.20}

\path[draw=drawColor,line width= 0.6pt,line join=round] ( 83.07, 29.33) --
	( 83.07, 32.08);

\path[draw=drawColor,line width= 0.6pt,line join=round] (105.43, 29.33) --
	(105.43, 32.08);

\path[draw=drawColor,line width= 0.6pt,line join=round] (127.79, 29.33) --
	(127.79, 32.08);

\path[draw=drawColor,line width= 0.6pt,line join=round] (150.15, 29.33) --
	(150.15, 32.08);

\path[draw=drawColor,line width= 0.6pt,line join=round] (172.52, 29.33) --
	(172.52, 32.08);
\end{scope}
\begin{scope}
\path[clip] (  0.00,  0.00) rectangle (216.81,216.81);
\definecolor{drawColor}{gray}{0.30}

\node[text=drawColor,anchor=base,inner sep=0pt, outer sep=0pt, scale=  0.88] at ( 83.07, 20.63) {2};

\node[text=drawColor,anchor=base,inner sep=0pt, outer sep=0pt, scale=  0.88] at (105.43, 20.63) {4};

\node[text=drawColor,anchor=base,inner sep=0pt, outer sep=0pt, scale=  0.88] at (127.79, 20.63) {6};

\node[text=drawColor,anchor=base,inner sep=0pt, outer sep=0pt, scale=  0.88] at (150.15, 20.63) {8};

\node[text=drawColor,anchor=base,inner sep=0pt, outer sep=0pt, scale=  0.88] at (172.52, 20.63) {10};
\end{scope}
\begin{scope}
\path[clip] (  0.00,  0.00) rectangle (216.81,216.81);
\definecolor{drawColor}{RGB}{0,0,0}

\node[text=drawColor,anchor=base,inner sep=0pt, outer sep=0pt, scale=  1.10] at (122.20,  7.62) {Number of peaks $P$};
\end{scope}
\begin{scope}
\path[clip] (  0.00,  0.00) rectangle (216.81,216.81);
\definecolor{drawColor}{RGB}{0,0,0}

\node[text=drawColor,rotate= 90.00,anchor=base,inner sep=0pt, outer sep=0pt, scale=  1.10] at ( 47.39,105.88) {Number of DP iterations};
\end{scope}
\begin{scope}
\path[clip] (  0.00,  0.00) rectangle (216.81,216.81);
\definecolor{drawColor}{RGB}{0,0,0}

\node[text=drawColor,anchor=base west,inner sep=0pt, outer sep=0pt, scale=  1.32] at ( 66.85,201.55) {Zoom to $P \leq 10$ peaks};

\node[text=drawColor,anchor=base west,inner sep=0pt, outer sep=0pt, scale=  1.32] at ( 66.85,187.30) {(linear scales)};
\end{scope}
\end{tikzpicture}
\end{minipage}
\vskip -0.5cm
\caption{\label{fig:variable-peaks} Comparison of time to compute optimal model with at most $P$ peaks using
  Segment Neighborhood (grey lines) and Optimal Partitioning with proposed sequential search (green
  dots). GFPOP with sequential search
  (Algorithm~\ref{algo:seq-search}) was used to compute optimal models
  with different numbers of peaks $P$, for two data sets with
  $N\approx 10^6$. \textbf{Left:} the number of iterations is
  linear $O(P)$ for Segment Neighborhood (grey line) but empirically
  $O(\log P)$ for Optimal Partitioning with sequential search (green
  dots). \textbf{Right:} Optimal Partitioning is empirically faster
  for computing models with $P>5$ peaks (10 segments); Segment
  Neighborhood is faster for smaller models.}
\end{figure}

For two data sets with $N\approx 10^6$ we therefore recorded the
empirical number of times GFPOP was called by the sequential search
algorithm. We observed that the number of GFPOP calls grows
logarithmically with $P$ (Figure~\ref{fig:variable-peaks}, left). For
a large number of peaks ($P>5$), it is clearly faster to use the
sequential search algorithm (Figure~\ref{fig:variable-peaks},
right). Overall these experiments indicate that the time complexity of
the sequential search in genomic data is $O(N\log(N)\log(P))$
for $N$ data and $P$ peaks.

\subsection{Application: computing a zero-error peak model}

\begin{figure}[t!]
\centering
\includegraphics{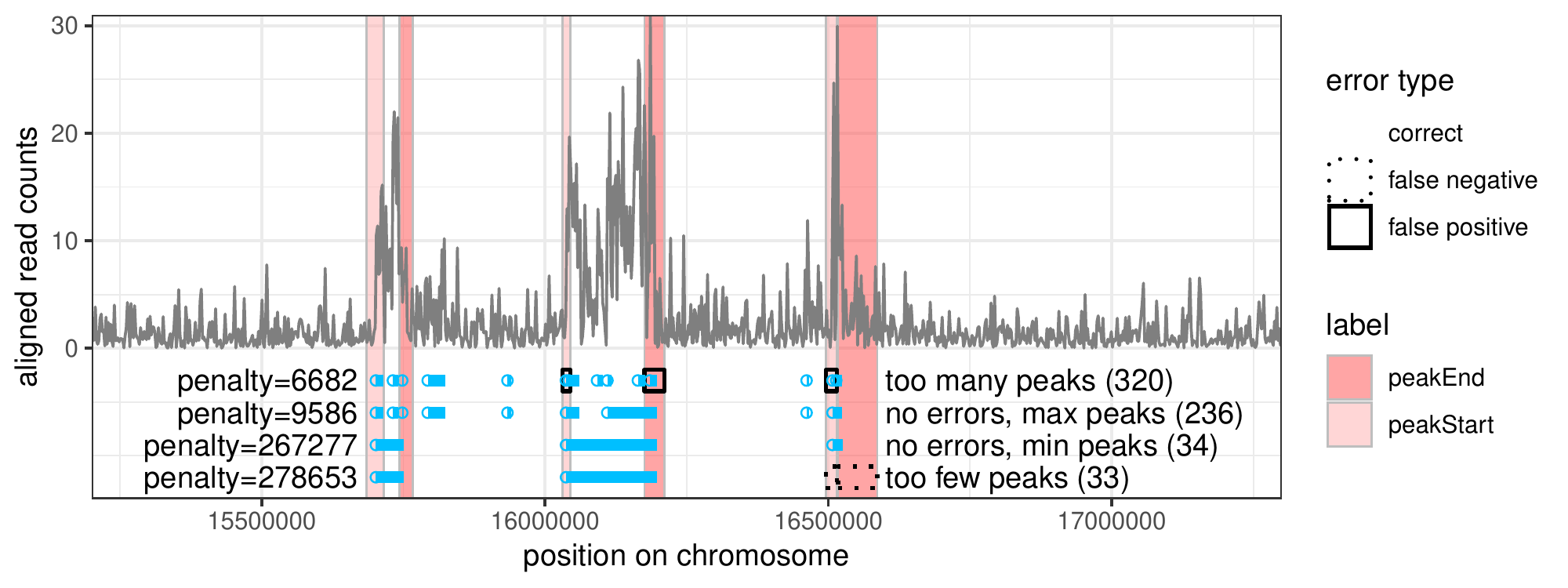}
\caption{\label{fig:label-error} Labels are used to compute an error
  rate for each peak model (blue bars), defined as the sum of false
  positive and false negative labels (rectangles with black
  outline). This H3K36me3 ChIP-seq data set has $N=1,254,751$ data to
  segment on a subset of chr12, but in the plot above we show only the
  82,233 data (grey signal) in the region around the labels (colored
  rectangles). The model with penalty=6682 results in 320 peaks, which
  is too many (three false positive labels with more than one peak
  start/end). Conversely, the model with penalty=278653 results in 33
  peaks, which is too few (only two peaks in the plotted region,
  resulting in two false negative labels on the right where there
  should be exactly one peak start/end). The range of penalties between 9586 and
  267277 results in models with between 34 and 236 peaks, and achieves
  zero label errors. }
\end{figure}

In this section we study the perfomance of the proposed algorithms in
a typical application. In the UCI \verb|chipseq| data set, there are
labels that indicate subsets of the data with or without peaks. In
this context the labels can be used to compute false positive and
false negative rates for any peak model. For example
Figure~\ref{fig:label-error} shows one data set with six labels and
four peak models computed via GFPOP. Small penalties result in too
many peaks, and large false positive rates. Large penalties result in
too few peaks, and large false negative rates. A range of intermediate
penalties/peaks achieves zero label errors. 
The labels can thus be used to determine an appropriate number of
peaks (with zero errors) for each data set.

\begin{figure}[t!]
\centering
\input{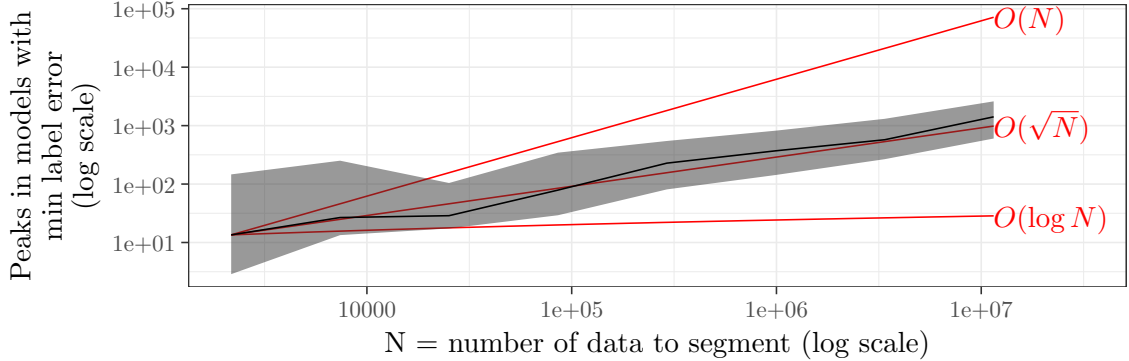}
\caption{\label{fig:data-peaks} The model with minimal label
  errors has $O(\sqrt{N})$ peaks in a data set of size $N$. For each data set we
  computed peak models with minimal label errors (see
  Figure~\ref{fig:label-error}); we then plot the number of peaks in
  minimal error models as a function of data set size $N$. Black median line
  and grey quartile band computed over several data sets of a given size
  $N$; asymptotic reference lines shown in red.}
\end{figure}

More generally, after computing GFPOP models for a range of penalties
for each data set, we computed the label error of each model. For each
data set we computed the min/max peaks that achieves zero label errors
(34/236 in Figure~\ref{fig:label-error}), along with the mean of those
two values, $(34+236)/2=135$. We plot the mean number of peaks that
achieves zero label errors as a function of data set size in
Figure~\ref{fig:data-peaks}. In these data it is clear that models
with $O(\sqrt N)$ peaks achieve zero label errors.

Computing the optimal model with $O(\sqrt N)$ peaks is computationally
expensive using the Segment Neighborhood algorithm
(\pkg{PeakSegOptimal} package), because the overall complexity would
be $O(N\sqrt N\log N)$. For example in $N=10^7$ data,
$P=O(\sqrt N)=1414$ peaks achieves zero label errors. Computing the
optimal model with Segment Neighborhood would thus require 2828
$O(N\log N)$ DP iterations. If we assume that each iteration would
have similar computational requirements as one $O(N\log N)$ run of
GFPOP, each would require about 1 hour and 80 gigabytes
(Figure~\ref{fig:intervals}). Overall that would mean 220 terabytes of
storage and 17 weeks of computation time, which is much too expensive
in practice.

\begin{figure}[t!]
\centering
\begin{minipage}{3in}
  \input{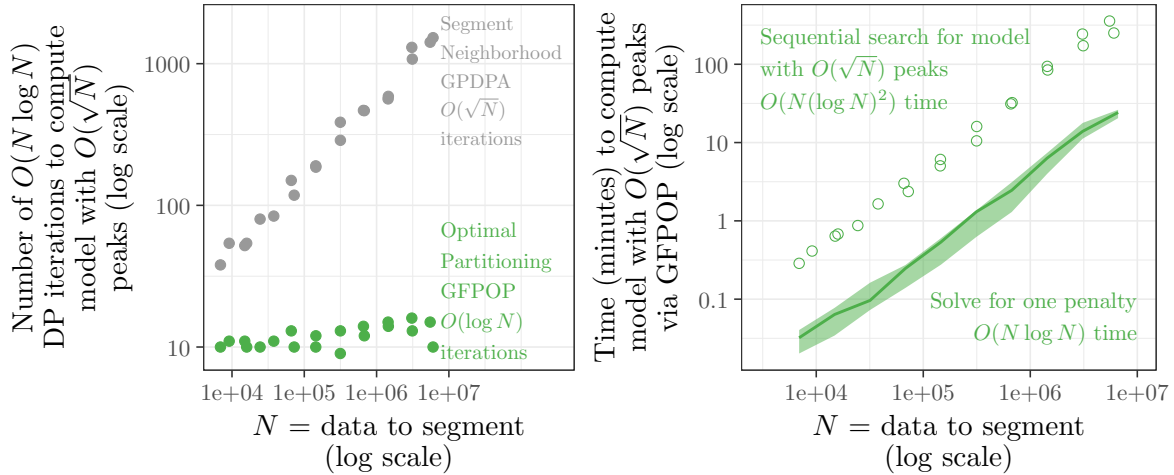}
\end{minipage} 
\begin{minipage}{3in} 
\begin{tikzpicture}[x=1pt,y=1pt]
\definecolor{fillColor}{RGB}{255,255,255}
\path[use as bounding box,fill=fillColor,fill opacity=0.00] (0,0) rectangle (224.04,187.90);
\begin{scope}
\path[clip] (  0.00,  0.00) rectangle (224.04,187.90);
\definecolor{drawColor}{RGB}{255,255,255}
\definecolor{fillColor}{RGB}{255,255,255}

\path[draw=drawColor,line width= 0.6pt,line join=round,line cap=round,fill=fillColor] (  0.00,  0.00) rectangle (224.04,187.90);
\end{scope}
\begin{scope}
\path[clip] ( 61.67, 43.96) rectangle (218.54,182.40);
\definecolor{fillColor}{RGB}{255,255,255}

\path[fill=fillColor] ( 61.67, 43.96) rectangle (218.54,182.40);
\definecolor{drawColor}{gray}{0.92}

\path[draw=drawColor,line width= 0.3pt,line join=round] ( 61.67, 56.01) --
	(218.54, 56.01);

\path[draw=drawColor,line width= 0.3pt,line join=round] ( 61.67, 85.65) --
	(218.54, 85.65);

\path[draw=drawColor,line width= 0.3pt,line join=round] ( 61.67,115.29) --
	(218.54,115.29);

\path[draw=drawColor,line width= 0.3pt,line join=round] ( 61.67,144.93) --
	(218.54,144.93);

\path[draw=drawColor,line width= 0.3pt,line join=round] ( 61.67,174.57) --
	(218.54,174.57);

\path[draw=drawColor,line width= 0.3pt,line join=round] ( 69.73, 43.96) --
	( 69.73,182.40);

\path[draw=drawColor,line width= 0.3pt,line join=round] (110.21, 43.96) --
	(110.21,182.40);

\path[draw=drawColor,line width= 0.3pt,line join=round] (150.69, 43.96) --
	(150.69,182.40);

\path[draw=drawColor,line width= 0.3pt,line join=round] (191.17, 43.96) --
	(191.17,182.40);

\path[draw=drawColor,line width= 0.6pt,line join=round] ( 61.67, 70.83) --
	(218.54, 70.83);

\path[draw=drawColor,line width= 0.6pt,line join=round] ( 61.67,100.47) --
	(218.54,100.47);

\path[draw=drawColor,line width= 0.6pt,line join=round] ( 61.67,130.11) --
	(218.54,130.11);

\path[draw=drawColor,line width= 0.6pt,line join=round] ( 61.67,159.75) --
	(218.54,159.75);

\path[draw=drawColor,line width= 0.6pt,line join=round] ( 89.97, 43.96) --
	( 89.97,182.40);

\path[draw=drawColor,line width= 0.6pt,line join=round] (130.45, 43.96) --
	(130.45,182.40);

\path[draw=drawColor,line width= 0.6pt,line join=round] (170.93, 43.96) --
	(170.93,182.40);

\path[draw=drawColor,line width= 0.6pt,line join=round] (211.41, 43.96) --
	(211.41,182.40);
\definecolor{drawColor}{RGB}{77,175,74}

\node[text=drawColor,anchor=base west,inner sep=0pt, outer sep=0pt, scale=  0.85] at ( 68.80,167.58) {Sequential search for model};

\node[text=drawColor,anchor=base west,inner sep=0pt, outer sep=0pt, scale=  0.85] at ( 68.80,155.29) {with $O(\sqrt N)$ peaks};

\node[text=drawColor,anchor=base west,inner sep=0pt, outer sep=0pt, scale=  0.85] at ( 68.80,143.00) {$O(N(\log N)^2)$ time};

\node[text=drawColor,anchor=base east,inner sep=0pt, outer sep=0pt, scale=  0.85] at (211.41, 67.63) {Solve for one penalty};

\node[text=drawColor,anchor=base east,inner sep=0pt, outer sep=0pt, scale=  0.85] at (211.41, 55.34) {$O(N \log N)$ time};
\definecolor{fillColor}{RGB}{77,175,74}

\path[fill=fillColor,fill opacity=0.50] ( 83.50, 59.15) --
	( 96.90, 67.53) --
	(110.31, 77.05) --
	(123.71, 83.76) --
	(137.12, 93.89) --
	(150.53,104.46) --
	(163.93,115.09) --
	(177.34,126.21) --
	(190.75,137.54) --
	(204.15,142.52) --
	(204.15,139.36) --
	(190.75,131.57) --
	(177.34,118.29) --
	(163.93,103.83) --
	(150.53, 94.39) --
	(137.12, 83.87) --
	(123.71, 75.09) --
	(110.31, 66.72) --
	( 96.90, 57.05) --
	( 83.50, 50.26) --
	cycle;

\path[draw=drawColor,line width= 1.1pt,line join=round] ( 83.50, 56.21) --
	( 96.90, 65.03) --
	(110.31, 70.34) --
	(123.71, 82.47) --
	(137.12, 92.42) --
	(150.53,103.88) --
	(163.93,112.04) --
	(177.34,124.25) --
	(190.75,134.36) --
	(204.15,141.50);

\path[draw=drawColor,line width= 0.4pt,line join=round,line cap=round] ( 83.50, 84.39) circle (  1.96);

\path[draw=drawColor,line width= 0.4pt,line join=round,line cap=round] ( 88.39, 89.05) circle (  1.96);

\path[draw=drawColor,line width= 0.4pt,line join=round,line cap=round] ( 97.12, 94.70) circle (  1.96);

\path[draw=drawColor,line width= 0.4pt,line join=round,line cap=round] ( 98.19, 95.48) circle (  1.96);

\path[draw=drawColor,line width= 0.4pt,line join=round,line cap=round] (105.69, 98.70) circle (  1.96);

\path[draw=drawColor,line width= 0.4pt,line join=round,line cap=round] (113.36,106.89) circle (  1.96);

\path[draw=drawColor,line width= 0.4pt,line join=round,line cap=round] (123.15,114.68) circle (  1.96);

\path[draw=drawColor,line width= 0.4pt,line join=round,line cap=round] (124.78,111.58) circle (  1.96);

\path[draw=drawColor,line width= 0.4pt,line join=round,line cap=round] (136.87,121.29) circle (  1.96);

\path[draw=drawColor,line width= 0.4pt,line join=round,line cap=round] (136.94,123.72) circle (  1.96);

\path[draw=drawColor,line width= 0.4pt,line join=round,line cap=round] (150.62,130.75) circle (  1.96);

\path[draw=drawColor,line width= 0.4pt,line join=round,line cap=round] (150.66,136.19) circle (  1.96);

\path[draw=drawColor,line width= 0.4pt,line join=round,line cap=round] (163.57,144.75) circle (  1.96);

\path[draw=drawColor,line width= 0.4pt,line join=round,line cap=round] (164.06,145.22) circle (  1.96);

\path[draw=drawColor,line width= 0.4pt,line join=round,line cap=round] (177.36,158.95) circle (  1.96);

\path[draw=drawColor,line width= 0.4pt,line join=round,line cap=round] (177.51,157.58) circle (  1.96);

\path[draw=drawColor,line width= 0.4pt,line join=round,line cap=round] (190.66,171.18) circle (  1.96);

\path[draw=drawColor,line width= 0.4pt,line join=round,line cap=round] (190.93,166.77) circle (  1.96);

\path[draw=drawColor,line width= 0.4pt,line join=round,line cap=round] (200.92,176.11) circle (  1.96);

\path[draw=drawColor,line width= 0.4pt,line join=round,line cap=round] (202.54,171.58) circle (  1.96);
\definecolor{drawColor}{gray}{0.20}

\path[draw=drawColor,line width= 0.6pt,line join=round,line cap=round] ( 61.67, 43.96) rectangle (218.54,182.40);
\end{scope}
\begin{scope}
\path[clip] (  0.00,  0.00) rectangle (224.04,187.90);
\definecolor{drawColor}{gray}{0.30}

\node[text=drawColor,anchor=base east,inner sep=0pt, outer sep=0pt, scale=  0.88] at ( 56.72, 67.58) {0.1};

\node[text=drawColor,anchor=base east,inner sep=0pt, outer sep=0pt, scale=  0.88] at ( 56.72, 97.22) {1};

\node[text=drawColor,anchor=base east,inner sep=0pt, outer sep=0pt, scale=  0.88] at ( 56.72,126.86) {10};

\node[text=drawColor,anchor=base east,inner sep=0pt, outer sep=0pt, scale=  0.88] at ( 56.72,156.50) {100};
\end{scope}
\begin{scope}
\path[clip] (  0.00,  0.00) rectangle (224.04,187.90);
\definecolor{drawColor}{gray}{0.20}

\path[draw=drawColor,line width= 0.6pt,line join=round] ( 58.92, 70.83) --
	( 61.67, 70.83);

\path[draw=drawColor,line width= 0.6pt,line join=round] ( 58.92,100.47) --
	( 61.67,100.47);

\path[draw=drawColor,line width= 0.6pt,line join=round] ( 58.92,130.11) --
	( 61.67,130.11);

\path[draw=drawColor,line width= 0.6pt,line join=round] ( 58.92,159.75) --
	( 61.67,159.75);
\end{scope}
\begin{scope}
\path[clip] (  0.00,  0.00) rectangle (224.04,187.90);
\definecolor{drawColor}{gray}{0.20}

\path[draw=drawColor,line width= 0.6pt,line join=round] ( 89.97, 41.21) --
	( 89.97, 43.96);

\path[draw=drawColor,line width= 0.6pt,line join=round] (130.45, 41.21) --
	(130.45, 43.96);

\path[draw=drawColor,line width= 0.6pt,line join=round] (170.93, 41.21) --
	(170.93, 43.96);

\path[draw=drawColor,line width= 0.6pt,line join=round] (211.41, 41.21) --
	(211.41, 43.96);
\end{scope}
\begin{scope}
\path[clip] (  0.00,  0.00) rectangle (224.04,187.90);
\definecolor{drawColor}{gray}{0.30}

\node[text=drawColor,anchor=base,inner sep=0pt, outer sep=0pt, scale=  0.88] at ( 89.97, 32.51) {1e+04};

\node[text=drawColor,anchor=base,inner sep=0pt, outer sep=0pt, scale=  0.88] at (130.45, 32.51) {1e+05};

\node[text=drawColor,anchor=base,inner sep=0pt, outer sep=0pt, scale=  0.88] at (170.93, 32.51) {1e+06};

\node[text=drawColor,anchor=base,inner sep=0pt, outer sep=0pt, scale=  0.88] at (211.41, 32.51) {1e+07};
\end{scope}
\begin{scope}
\path[clip] (  0.00,  0.00) rectangle (224.04,187.90);
\definecolor{drawColor}{RGB}{0,0,0}

\node[text=drawColor,anchor=base,inner sep=0pt, outer sep=0pt, scale=  1.10] at (140.10, 19.50) {$N$ = data to segment};

\node[text=drawColor,anchor=base,inner sep=0pt, outer sep=0pt, scale=  1.10] at (140.10,  7.62) {(log scale)};
\end{scope}
\begin{scope}
\path[clip] (  0.00,  0.00) rectangle (224.04,187.90);
\definecolor{drawColor}{RGB}{0,0,0}

\node[text=drawColor,rotate= 90.00,anchor=base,inner sep=0pt, outer sep=0pt, scale=  1.10] at ( 13.63,113.18) {Time (minutes) to compute};

\node[text=drawColor,rotate= 90.00,anchor=base,inner sep=0pt, outer sep=0pt, scale=  1.10] at ( 25.51,113.18) {     model with $O(\sqrt N)$ peaks};

\node[text=drawColor,rotate= 90.00,anchor=base,inner sep=0pt, outer sep=0pt, scale=  1.10] at ( 37.39,113.18) {     via GFPOP (log scale)};
\end{scope}
\end{tikzpicture} 
\end{minipage} 
\vskip -0.5cm
\caption{\label{fig:evaluations} Computing a zero-error model with
  $O(\sqrt{N})$ peaks is possible in $O(N(\log N)^2)$ time using our
  proposed Optimal Partitioning Search algorithm. \textbf{Left:}
  Segment Neighborhood requires $O(\sqrt{N})$ dynamic programming
  iterations to compute a model with $O(\sqrt{N})$ peaks; our proposed
  Optimal Partitioning search algorithm requires only $O(\log N)$
  iterations. \textbf{Right:} Optimal Partitioning solves for one
  penalty in $O(N\log N)$ space/time (median line and 5\%/95\% quantile band
  over data sets and penalties); finding the zero-error model with
  $O(\sqrt{N})$ peaks takes $O(N (\log N)^2)$ time/space -- only a log
  factor more (points).}
\end{figure}

Instead, we can use the proposed sequential search
(Algorithm~\ref{algo:seq-search}) to compute a zero-error model with
$O(\sqrt N)$ peaks. In our empirical tests, we observed that only
$O(\log N)$ GFPOP calls are required to compute $O(\sqrt N)$ peaks
(Figure~\ref{fig:evaluations}, left). In particular for $N=10^7$ data
only 10--15 GFPOP calls are required, which is significantly fewer
than the 2828 DP iterations that would be required for the Segment
Neighborhood solver in the \pkg{PeakSegOptimal} package. 

We also observed that the empirical timings of the sequential search
are only a log-factor slower than solving for one penalty
(Figure~\ref{fig:evaluations}, right). In particular for $N=10^7$ data
the sequential search takes on the order of hours, which is much less
than the weeks that would be required to solve the Segment
Neighborhood problem. Overall these empirical results indicate that
the sequential search algorithm in the \pkg{PeakSegDisk} package
can be used to compute a model with $O(\sqrt N)$ peaks in
$O(N(\log N)^2)$ time.

\section{Summary and discussion} \label{sec:summary} 

This paper presented two new algorithms for constrained optimal
changepoint detection. We presented Generalized Functional
Partitioning Optimal Partitioning (GFPOP) which computes the optimal
model for one penalty $\lambda$. We also proposed a sequential search
algorithm which repeated calls GFPOP in order to compute the most
likely model with at most $P$ peaks. 

We analyzed the proposed algorithms by running them on a set of
genomic data sets ranging from $N=10^3$ to $N=10^7$. First, we showed
that the algorithms can be used to compute models which are more
likely than existing heuristics, and often simpler (fewer
peaks). 

Second, we studied the empirical complexity of GFPOP as the a function
of the number of data $N$. We showed that GFPOP requires $O(N\log N)$
time, $O(N \log N)$ space, and $O(\log N)$ memory. We furthermore
showed that using disk-based storage is only a constant factor slower
than memory-based storage. Overall we showed that GFPOP can be used to
compute optimal peak models for up to $N=10^7$ data in reasonable
amounts of time (minutes). 

Third, we studied the empirical complexity of sequential search as a
function of data size $N$ and number of peaks $P$. We showed that it
requires $O(N\log(N)\log(P))$ time, $O(N\log N)$ disk, $O(\log N)$
memory. In particular we showed that it is always faster than Segment
Neighborhood solvers for models with $P>5$ peaks. 

Finally we analyzed the labels in our benchmark of genomic data,
which indicated that an appropriate number of peaks is $P=O(\sqrt
N)$. We showed that sequential search computes the model with
$O(\sqrt N)$ peaks in $O(N(\log N)^2)$ time, whereas existing Segment
Neighborhood algorithms would be $O(N\sqrt N\log N)$. We showed that
for $N=10^7$ data our approach only requires hours/gigabytes of
time/space to compute optimal models. Our algorithms thus make it
practical for the first time to compute optimal models with many peaks
for genomic data sets.

\paragraph{Reproducible research statement.} The source code and data used to create this manuscript (including all figures) is available at \url{https://github.com/tdhock/PeakSegFPOP-paper}







\bibliographystyle{abbrvnat}
\bibliography{jss-refs}

\end{document}